\documentclass[pra,twocolumn,showkeywords,longbibliography,superscriptaddress,10pt]{revtex4-1}
\usepackage{appendix}
\usepackage{graphicx}
\usepackage{dcolumn}
\usepackage{color}
\usepackage{times}
\usepackage{bm}
\usepackage{amssymb}
\usepackage{amsmath}
\usepackage{epsfig}
\usepackage{epstopdf}
\usepackage{dsfont}
\usepackage{subfigure}
\usepackage{tikz}
\usepackage[colorlinks, citecolor=blue, linkcolor=blue,urlcolor=blue]{hyperref}
\usepackage[mathscr]{euscript}
\makeatletter
\renewcommand{\maketag@@@}[1]{\hbox{\m@th\normalsize\normalfont#1}}
\makeatother
\begin{document}
	\renewcommand{\thefootnote}{\fnsymbol {footnote}}
	
	 \title{Showcasing extractable work of quantum battery via entropic uncertainty relations}
	%Effect of Hawking radiation on entropic uncertainty relations in the presence of quantum memory
	%
	\author{Meng-Long Song}
	\affiliation{School of Physics \& Optoelectronic Engineering, Anhui University, Hefei
		230601,  People's Republic of China}
	
	%\author{Li-Juan Li}
%	\affiliation{School of Physics \& Optoelectronic  Engineering, Anhui University, Hefei 230601, People's Republic of China}
	
	\author{Xue-Ke Song}
	\affiliation{School of Physics \& Optoelectronic Engineering, Anhui University, Hefei 230601,  People's Republic of China}
	
	\author{Liu Ye}
	\affiliation{School of Physics \& Optoelectronic Engineering, Anhui University, Hefei 230601,  People's Republic of China}
	
	\author{Dong Wang} \email{dwang@ahu.edu.cn}
	\affiliation{School of Physics \& Optoelectronic Engineering, Anhui University, Hefei
		230601,  People's Republic of China}

	\date{\today}
	
	\begin{abstract}
In this study, we investigate the effectiveness of entropic uncertainty relations (EURs) in discerning the energy variation in quantum batteries (QBs) modelled by battery-charger-field in the presence of bosonic and fermionic reservoirs. Our results suggest that the extractable works (exergy and ergotropy) have versatile characteristics in different scenarios, resulting in a complex relationship between tightness and extractable work. It is worth noting that the tightness of the lower bound of entropic uncertainty can be a good indicator for energy conversion efficiency in charging QBs. Furthermore, we disclose how the EUR including uncertainty and lower bound contributes to energy conversion efficiency in the QB system.
It is believed that these findings will be beneficial for better understanding the role of quantum uncertainty in evaluating quantum battery performance.
%In essence, the energy transfer of the charging system can  illustrate the performance of quantum batteries (QBs), and thus to reveal the features of which are of basic importance in the frame of QBs. We here take into account the charger-battery entropic uncertainty relation to  insight into the energy transfer of the charging protocol. We found that the tighter entropic bound promotes energy storage, including the extractable work (exergy or ergotropy), when the bosonic reservoir  interacts to the charge-battery system. In particular, if the charger-battery system is considered to be detuning with the driving field, the strong  tightness will be in favor of all extractable work (both exergy and ergotropy), as well as for the fermion reservoir. The mechanism that the entropic uncertainty relation contributes to the energy storage and extraction  is  revealed in detail, and the results show that both the uncertainty and its lower bound are capable of directly indicating the variation of energy storage and exergy. Finally, it is found that the magnitude of the uncertainty dominates the  ergotropy. Thus, we argue that our findings
%provide new avenue to probe the extractable work in QB system.
	\end{abstract}
	
	\maketitle

	\section{Introduction}
	{The storage and utilisation of energy have always been an important topic worldwide and can be achieved by devices, that is,  so-called batteries. In contrast to the conversion of chemical energy into electric energy in the classical battery scheme, quantum batteries (QBs) \cite{review} are regarded as a new and promising energy storage process, pioneeringly presented by Alicki and Fannes \cite{propose}. The core principle of quantum batteries is using quantum resources (e.g., quantum entanglement and coherence) for energy storage \cite{resource1,resource2,resource3}. With the development of quantum information theory, quantum batteries have received considerable attention, and progress has been made in theory, including the construction of charging models \cite{model1,model2,model3,model4,model5,model6,model7},
open QB systems \cite{open1,open2,open3,open4,open5,open6,open7,open8,Dou-F-Q1}, many-body QBs (e.g., Dicke quantum battery and spin-based batteries) \cite{model4,r1,cells1,cells2,cells3,cells4,Dou-F-Q2},
and so on. Several promising experiments have been conducted for QBs in organic microcavities \cite{experiment1}, superconducting \cite{experiment2}, quantum dots \cite{experiment4} and NMR \cite{experiment3} platforms.

The great contribution of quantum resources is to allow QBs to have superextensive scaling of the charging power in many-body QB systems, such as entanglement of battery subspace resulting in quantum advantages over classical schemes \cite{review,model1,model2}. However, many investigations \cite{review,open8,Dou-F-Q2,x18,x18++,capacity,measurement} have revealed that quantum sources cannot independently and adequately predict the energy variation in QBs. This is due to several reasons, such as non-equilibrium thermodynamic processes in which multiple quantum resources alternately or simultaneously act on the battery. Therefore, it is argued that the analysis of QB's performance needs more indicators in the specific charging protocol.
Conversely, G$\rm \ddot{u}$hne and Lewenstein have studied the intrinsic relationship between entanglement and entropic uncertainty relations (EURs) \cite{ee}.
With this in mind, it is natural to ask whether EURs can be used to predict the change in quantum battery's energy. Upon studying the relationship between QB's energy and EURs, we found the answer to be positive.
%In the common charging scheme, the battery and the energy source are mediated by the charger, and the charger is coupled with the energy source and the battery to transfer energy simultaneously. This process is usually accompanied by the system-energy loss because of unavoidably coupling with its surrounding environment.  Actually, there are several crucial factors to  evaluate  QBs' performance as: (i) one is the maximum average charging power, which relates to capture the most energy in the shortest time; (ii) another is regarded as the ability to completely transfer the energy stored in the battery to the extractor, which quantifies capability the energy storage of extracting the useful work \cite{model2,model4,reference4,reference5}, and (iii) the other is the efficiency of energy transfer, being related to the cost during charging. Incidentally,
%with respect to a steady-state QB, the performance of QBs relies on amount of the most energy stored and extraction of all of the energy.

The uncertainty relation, originally proposed by Heisenberg, is the cornerstone of quantum information theory. It proposes that one cannot accurately predict the measurement outcomes with respect to two arbitrary incompatible observables \cite{1}. Afterwards, Kennard \cite{2} and Robertson \cite{3} generalized into the inequality for arbitrary two operators: $\Delta {\hat{M}_1} \cdot \Delta {\hat{M}_2} \ge | \langle {[{\hat{M}_1},{\hat{M}_2}]}  \rangle |/2$ in terms of standard deviation, with $\Delta (\bullet)$ denoting the standard deviation and $ \langle \circ\rangle$ denoting the expected value. Eversince, there have been various quantifications of uncertainty relations, including Shannon entropy \cite{4}, R\'{e}nyi entropy \cite{6}, and von Neumann entropy \cite{JR}. Furthermore, Berta {\it et al.} \cite{7} presented quantum-memory-assisted entropic uncertainty relations (QMA-EURs).
Liu $et$ $al.$ novelly derived an expression for QMA-EURs as multiple observables \cite{liu}
		\begin{align}
			\sum\limits_{x = 1}^m {S\left( {{{\hat M}_x}\left| B \right.} \right)}  \ge   {\log _2}\frac1f + \left( {m - 1} \right)S\left( {A\left| B \right.} \right),
			\label{Eq.1}
		\end{align}
herein, the von Neumann conditional entropy $S ( {A|B} ) = S ( {{{\hat \rho }_{AB}}}  ) - S\left( {{{\hat \rho }_B}} \right)$  \cite{m,x,wu-lin} with $S\left( {\hat \rho } \right) =  - tr\left( {\hat \rho {{\log }_2}\hat \rho } \right)$, and $f= \mathop {\max }\limits_{{i_N}} \left\{ {\sum\limits_{{i_2} \sim {i_{n - 1}}} {\mathop {\max }\limits_{{i_1}} } \left[ {c\left( {u_{{i_1}}^1,u_{{i_2}}^2} \right)} \right]\prod\limits_{x = 2}^{n- 1} {\left[ {c\left( {u_{{i_x}}^x,u_{{i_{x + 1}}}^{x + 1}} \right)} \right]} } \right\}$, where $u_{i_x}^x$ is the $i$th eigenvector of the operator ${{{\hat M}_x}}$ and ${c\left( {u_{{i_x}}^x,u_{{i_{x + 1}}}^{x + 1}} \right)}$ denotes the maximal overlap of the operators ${u_{{i_x}}^x}$ and $u_{{i_{x + 1}}}^{x + 1}$.

In this study, we focused on the effectiveness of EURs in discerning the energy variation for QBs modelled by a charger-mediated protocol with an external field and reservoir. We explored the relationship between the extractable work and the EUR of the QB. Compared with previous results \cite{reference1,x18,x18++,capacity}, the merits of the current work can be summarised as follows: (i) Improving the extractable work (i.e., exergy and ergotropy) and conversion efficiency of the battery was revealed. (ii) We found that EUR tightness could serve as an excellent indictor for predicting extractable work and conversion efficiency.
(iii) We revealed how the corresponding EURs contribute to the conversion efficiency of QB.
%That is, the lower bound and conditional entropy $S\left[ {\hat \rho \left( {{{\hat \sigma }_z}\left| B \right.} \right)} \right]$ are synchronized with the internal energy and exergy of the battery, while the conditional entropy $S\left[ {\hat \rho \left( {{{\hat \sigma }_x}\left| B \right.} \right)} \right]$ is responsible for ergotropy. }
%In general, entanglement is regarded as a crucial role in energy transfer of QBs \cite{x18,x18++}. Besides,  G$\rm \ddot{u}$hne and  Lewenstein have studied the intrinsic relationship  between entanglement and EURs \cite{ee}.
%Considering these together, it is an interesting topic whether there is any intrinsic relationship between QBs' extractable work and EURs.
%Inspired by this, we first investigate the internal energy and extractable work in the paradigmatic model of battery-charger-field with bosonic and fermion reservoirs, which makes use of EURs to capture better energy storage, extractable work and extraction efficiency in QBs. Remarkably, it is  interesting to find that the tightness of EURs perfectly indicates quantum batteries'  extractable work. Thus, we claim that the tightness of the uncertainty relation is available for indicating the energy extraction in practical quantum batteries.

The remainder of this paper is organised as follows. Section II introduces a typical quantum-battery model with a battery-charger-field in the presence of reservoirs. In Section III, in addition to the performance of the QB, the relationship between extractable work and tightness is investigated. In Section IV, the mechanism for improving extractable work via EUR is provided. Finally, our article ends with a brief discussion and conclusions in Section V.

\section{The model of quantum battery}
The diagrammatic charging model of a QB consists of a battery and a two-level charger. Simultaneously, the charger is powered by a driving laser field and coupled to a bosonic or fermionic reservoir. As illustrated in Fig. \ref{f1}, the energy can be transferred to the battery owing to the interaction between the charger and the battery. Consequently, the energy storage of the battery was achieved. In the rotating frame, the Hamiltonian of the charging system
 ${{\cal H}_s}$ can be written as \cite{reference1,reference2,reference3}
	\begin{figure}
		\begin{minipage}{0.45\textwidth}
			\centering
			\subfigure{\includegraphics[width=8cm]{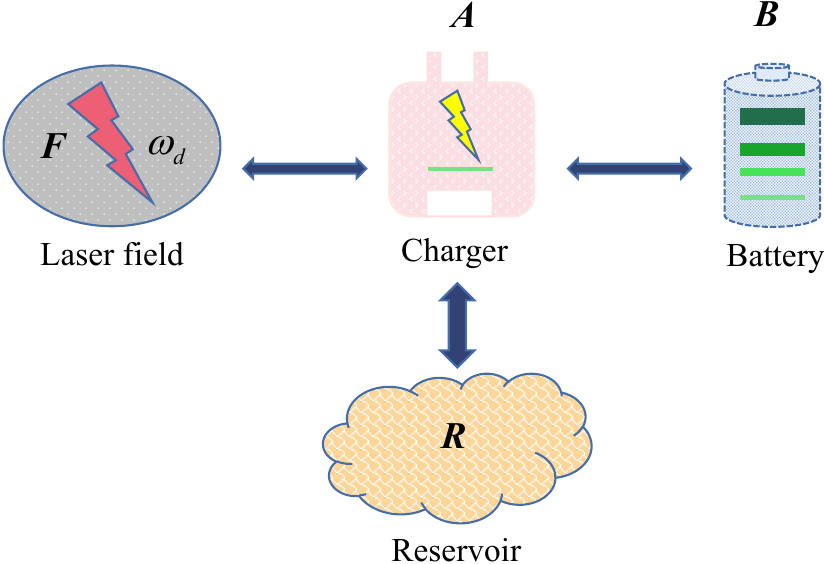}}
		\end{minipage}\hfill
		\caption{Schematic diagram of the charging process of the quantum battery. The charger ($A$) is driven by a classical laser field with frequency ${\omega _d}$ and amplitude $F$,
and is coupled with the battery ($B$) and interacts with its surrounding reservoir ($R$). As a result, the energy is not only transferred from the charger to the battery ($B$) but is also dissipated into the reservoir ($R$) mediated by the charger. The two-level charger and the battery are characterized by transition frequencies ${\omega _A}$ and ${\omega _B}$, respectively.}
		\label{f1}
	\end{figure}
\begin{align}
       {{\cal H}_s}= &\frac{{{\Delta _A}}}{2}\hat{\sigma} _z^A + \frac{{{\Delta _B}}}{2}\hat{\sigma} _z^B+  {g}\left( {\hat{\sigma} _ + ^A\hat{\sigma} _ - ^B + \hat{\sigma} _ - ^A\hat{\sigma} _ + ^B} \right) \nonumber\\
            & + \frac{F}{2}\left( {\hat{\sigma} _ + ^A + \hat{\sigma} _ - ^A} \right),
		\label{Eq.2}
	\end{align}
here, ${{\cal H}_s}$ denotes the Hamiltonian of the charger-battery system (the Boltzmann constant ${{k_B} =\hbar  = 1}$ hereafter, for simplicity).
%${{\cal H}_r}$ represents the Hamiltonian of the reservoir, and  $\cal V$ denotes the Hamiltonian for the charger-reservoir coupling.
Moreover, $\hat{\sigma} _j^{A,B}\left( {j = x,y,z } \right)$ are the Pauli operators and  $\hat{\sigma} _k^{A,B}\left( {k = + , - } \right)$ are the upper and down operators, ${\Delta _l} = {\omega _l} - {\omega _d}\left( {l = A,B} \right)$ is the detuning of a two-level system with a transition frequency of ${\omega _l}$ to the driving field with frequency ${\omega _d}$, assume the charger has the same frequency as the battery, i.e., ${\omega _A} = {\omega _B} = {\omega _0}$. ${g}$ is the coupling strength between the charger and battery, and $F$ is the amplitude of the driving field.

The Markovian master equation describing the charger-battery system can be obtained  \cite{experiment1,reference1,reference2,measurement,reference3}
\begin{align}
{\dot {\hat {\rho}} _{AB}} &=  - i\left[ {{{\cal H}_s},{\hat{\rho} _{AB}}} \right]\nonumber\\
 &+ J\left\{ {N\left( T  \right){D_{\hat{\sigma} _ - ^A}}\left[ {{\hat{\rho} _{AB}}} \right] + n\left( T  \right){D_{\hat{\sigma} _ + ^A}}\left[ {{\hat{\rho} _{AB}}} \right]} \right\},
		\label{Eq.3}
	\end{align}
	where ${D_\Lambda}\left[ \hat{\rho}  \right] = 2\Lambda\hat{\rho} {\Lambda^\dag } - {\Lambda^\dag }\Lambda\hat{\rho}  - \hat{\rho} {\Lambda^\dag }\Lambda$ denotes the dissipator, and $J$ is the dissipative rate. Regarding the different types of reservoirs, the average particle number $n(T)$ is different at frequency ${\omega_k }$ and temperature $T$. Specifically, $n\left( T  \right) = {n_b} = 1/\left[ {\exp \left( {\omega_k /T} \right) - 1} \right] $ and $N\left( T  \right) = 1 + n\left( T  \right)$ in the bosonic reservoir. For the fermionic one, $n\left( T  \right) = {n_f} = 1/\left[ {\exp \left( {\omega_k  - \mu } \right)/T + 1} \right]$ and $N\left( T  \right) = 1 - n\left( T  \right)$, where  $\mu $ denotes the chemical potential of the reservoir.

Suppose that the initial state of the charger-battery is prepared in the ground state $\left| g \right\rangle $ (henceforth, the excited state is written as $\left| e \right\rangle $), that is,
	\begin{align}
		{\hat{\rho} _{AB}}\left( {t \le 0} \right) = {\left| {gg} \right\rangle _{AB}}\left\langle {gg} \right|.
		\label{Eq.4}
	\end{align}
Once the charging process begins $\left( {t \ge 0} \right)$, the driving field continues to power the charger, and the charger is connected to the reservoir because it interacts with the charger, whereas energy is transferred through the charger to the battery by interaction. The internal energy of the charger ($A$) or battery ($B$) at any time $t$ can be expressed as
\begin{align}
		{E_i}\left( t \right) = {\rm tr}\left[ {{{\cal H}_0^i}{\hat{\rho} _i}\left( t \right)} \right],
		\label{Eq.5}
	\end{align}
where ${{\cal H}_0^i} = {\omega_0} {\hat{\sigma} _ +^i }{\hat{\sigma} _ -^i }$ and ${{\hat{\rho} _i}\left( t \right)}$ ($i=A$ or $B$) is the reduced density matrix of ${{\hat{\rho} _{AB}}\left( t \right)}$. In addition to the storage capacity of the QB, the extractable work is also an important factor in determining the QB's performance. First, the extraction of useful work from a battery by thermalization (a nonunitary extraction process) can be quantified by the variations in the Helmholtz free energy of the battery. In other words, the battery is connected to a thermal bath with an inverse temperature $\beta$. Consequently, the work can be extracted as exergy \cite{x18++,exergy1,exergy2}
\begin{align}
		{W_f}\left( t \right) = \mathcal{F}\left( {{{\hat \rho }_B}} \right) - \mathcal{F}\left( {{{\hat \rho }_\beta }} \right),
		\label{Eq.6}
	\end{align}
where $\mathcal{F}\left( {\hat \rho_B } \right) = {\rm tr}\left( {{\mathcal{H}_0}\hat \rho_B } \right) - S\left( {\hat \rho_B } \right)/\beta $ is Helmholtz free energy, and ${{\hat \rho }_\beta } = {e^{ - \beta {\mathcal{H}_0^B}}}/{\rm tr}\left( {{e^{ - \beta {\mathcal{H}_0^B}}}} \right)$ is the thermal equilibrium state of the battery.
To characterize the maximal amount of energy that can be extracted from a QB in the charging process under the cyclic unitary operations, the ergotropy \cite{reference4,reference5} is introduced as ${W_e \left( t \right)} = {\rm tr}\left[ {{{\cal H}_0^B}{\hat{\rho} _B}\left( t \right)} \right] - {\rm tr}\left[ {{{\cal H}_0^B}{\hat{\rho} _{p}}} \right]$ where ${{\hat{\rho} _p}}$ is called the passive state of ${{\hat{\rho} _B}}(t)$ with the zero extractable energy by cyclic unitary operations. In spectral decomposition, the Hamiltonian $H$ and density matrix $\hat{\rho} $ are written as $H = \sum\nolimits_j {{e_j}} \left| {{e_j}} \right\rangle \left\langle {{e_j}} \right|$ and $\hat{\rho}  = \sum\nolimits_j {{r_j}} \left| {{r_j}} \right\rangle \left\langle {{r_j}} \right|$ respectively, where ${{r_j}}$ and ${{e_j}}$ are the eigenvalues of $\hat{\rho} $ and $H$ with corresponding eigenstates $\left| {{r_j}} \right\rangle $ and $\left| {{e_j}} \right\rangle $. After ordering eigenvalues (${r_0} \ge {r_1} \ge {r_2} \ge  \cdots $ and ${e_0} \le {e_1} \le {e_2} \cdots $), the passive state can be derived as ${\hat{\rho} _p} = \sum\nolimits_j {{r_j}} \left| {{e_j}} \right\rangle \left\langle {{e_j}} \right|$. The ergotropy can be reexpressed as
	%where, $\lambda $ is the spectrum width and ${1 \mathord{\left/
%			{\vphantom {1 \lambda }} \right.
%			\kern-\varthetalldelimiterspace} \lambda }$ is the correlation time of the reservoir, and $\vartheta $ is the effective coupling strength satisfying $L = \vartheta {\beta _T}$ where $L$ is the vacuum Rabi frequency. Then we define a dimensionless real number ${R} = L/{\lambda }$, where strong coupling has $ {{R} \gg 1} $ and weak coupling with ${{R} \ll 1} $ \cite{b191,b192,b193,b194,b195,b196,b197}. In this case, Markovian evolution occurs under weak coupling, and non-Markovian dynamics and memory effects occur under strong coupling. And then the analytic solution for ${\eta _i}\left( t \right) \left( {i = 1,2} \right)$ comes in the following form \cite{b23,b24,b}
\begin{align}
		{W_e \left( t \right)} = {E_B}\left( t \right) - \sum\limits_j {{r_j}} {e_j}.
		\label{Eq.7}
	\end{align}
Furthermore, combining the density matrix of the charger-battery system ${\hat{\rho} _{AB}}\left( t \right)$, ${E_B}\left( t \right)$ and $W\left( t \right)$ can be described as follows:
	\begin{align}
{E_B}\left( t \right) &= {\omega_0} \left[ {{\rho _{11}}\left( t \right) + {\rho _{33}}\left( t \right)} \right], \nonumber \\
\label{Eq.8}
\end{align}
and
\begin{align}
{W_e\left( t \right)} &= \frac{{\omega_0} }{2}\left\{ {\sqrt {4{{\left| {{\rho _{12}}\left( t \right) + {\rho _{34}}\left( t \right)} \right|}^2} + {\kappa ^2}}  + \kappa } \right\},
\label{Eq.9}
\end{align}
where ${\rho _{ij}}\left( t \right)\left( {i,j = 1,2,3,4} \right)$ is the $i$-th row and $j$-th column matrix entry of ${\hat{\rho} _{AB}}\left( t \right)$ with $\kappa  = 2\left[ {{\rho _{11}}\left( t \right) + {\rho _{33}}\left( t \right)} \right] - 1$. Note that when charging for a sufficiently long period (i.e., $t \to \infty $), the charger-battery system becomes steady, and the system's energy does not vary over time $t$. Hence, we have ${E_i}\left( t \right) \to {E_i}\left( \infty  \right)$, ${W_f \left( t \right)} \to {W_f \left( \infty  \right)}$ and ${W_e \left( t \right)} \to {W_e\left( \infty  \right)}$ (${\dot {\hat {\rho}} _{AB}} = 0$) in the limit.
In fact, when $S\left( {{{\hat \rho }_\beta }} \right) - S\left( {{{\hat \rho }_p}} \right) > \beta \left[ {{\rm tr}\left( {{\mathcal{H}_0^B}{{\hat \rho }_\beta }} \right) - {\rm tr}\left( {{\mathcal{H}_0^B}{{\hat \rho }_p}} \right)} \right]$, the battery  always stores more exergy than ergotropy. The left of the above inequality is the entropy production of the battery during the thermalization process, and the right is the change in battery energy caused by thermalization \cite{exergy2,exergy3}. In addition, the   work that can be extracted ${W_f \left( t \right)} \in \left[ {0,\Delta {E_B}\left( t \right)} \right]$ is hold, where $\Delta {E_B}\left( t \right) = {E_B}\left( t \right) - {E_B}\left(0 \right)$  represents the stored energy of the battery.

As a matter of fact, the conversion rate of energy is termed as another nontrivial factor to evaluate the performance of QB, in addition to the amount of energy. Therefore, the conversion efficiency of the battery is given by
\begin{align}
{\eta_1}\left( t \right) =\frac{{{W_f\left( t \right)}}}{{\Delta {E_B}\left( t \right)}},\nonumber \\
{\eta_2 }\left( t \right) =\frac{{{W_e \left( t \right)}}}{{\Delta {E_B}\left( t \right)}}.
\label{Eq.10}
\end{align}
It quantifies the proportion of useful work in QB's energy storage and deals with the energy cost \cite{review,open1,cost3} of charging QBs (higher efficiency means lower energy costs). Therefore, a high-performance QB essentially needs to its higher efficiency  to pursue more extractable work.
%As a matter of fact,
%the transfer rate of energy is termed as another nontrivial factor to evaluate the performance of QB, in addition to the amount of energy.
%Therefore, the extraction efficiency of the battery is given by
%\begin{align}
%{\eta }\left( t \right) =\frac{{W\left( t \right)}}{{\Delta {E_B}\left( t \right)}}.
%\label{Eq.12}
%\end{align}

	\section{Performance of quantum batteries}
In general, quantum batteries with excellent performances have two significant advantages: stability and high power. In this consideration, we investigated the energy changes for the considered charging scenarios while the state of the charger-battery system was steady. Importantly,
the role of the uncertainty's tightness of QMA-EUR is discussed in detail while determining the acquirement and extraction of QB's energy.

To probe the dynamics of entropic uncertainty, we use Pauli operators as the incompatible observables to be measured, which leads to post-measurement states $\left( {i = x,y,z} \right)$
	\begin{align}
		\hat{\rho} \left( {{\hat{\sigma} _i}\left| B \right.} \right) = \frac{1}{2}\left[ {{\hat{\rho} _{AB}}\left( t \right) + \left( {{\hat{\sigma} _i} \otimes \mathbb{I}} \right){\hat{\rho} _{AB}}\left( t \right)\left( {{\hat{\sigma} _i} \otimes \mathbb{I}} \right)} \right],
		\label{Eq.11}
	\end{align}
where $\mathbb{I}$ denotes an identity matrix. It can be observed that different measurement choices for $A$ lead to different post-measurement states. Subsequently, we explored the entropic uncertainty relationship of the system under different measurement choices. When two measurement operators $\left( {{\hat{\sigma} _x},{\hat{\sigma} _z}} \right)$ are selected, the entropic uncertainty (i.e., left-hand side of Eq.  (\ref{Eq.1})) is expressed as
\begin{align}
		U_l^{xz}\left( t \right) =\sum\limits_{j = x,z} {S\left[ {\hat \rho \left( {{{\hat \sigma }_j}\left| B \right.} \right)} \right]}  - 2S\left[ {{{\hat \rho }_B}\left( t \right)} \right],
		\label{Eq.12}
	\end{align}
and the corresponding lower bound (right-hand side of Eq. ~ (\ref{Eq.1})) can be expressed as
		\begin{align}
			U_r^{xz}\left( t \right) = S\left[ {{\hat{\rho} _{AB}}\left( t \right)} \right] - S\left[ {{\hat{\rho} _B}\left( t \right)} \right] + 1.
			\label{Eq.13}
		\end{align}
Consequently, the uncertainty's tightness is
\begin{align}
		{\Delta ^{xz}}\left( t \right) = U_l^{xz}\left( t \right) - U_r^{xz}\left( t \right).
		\label{Eq.14}
	\end{align}
Incidentally, a smaller value of ${\Delta ^{xz}}\left( t \right)$, i.e., stronger tightness-implies a tighter lower bound, and vice versa.
For the steady state,  $U_l^{xz}\left( t \right) \to U_l^{xz}\left( \infty  \right)$ and $U_r^{xz}\left( t \right) \to U_r^{xz}\left( \infty  \right)$; thus, we have ${\Delta ^{xz}}\left( t \right) \to {\Delta ^{xz}}\left( \infty  \right)$ in terms of Eq. (\ref{Eq.14}).

\subsection{Resonance}

To probe the energy-transfer issues of QBs, we first consider the case of eliminating detuning between the charger (battery) and driving field regarding the system's energy variation, that is, ${\Delta _A} = {\Delta _B} = 0$. A zero-temperature reservoir where $T \to 0$ leads to ${n_b}, {n_f}\to 0$. Consequently, the state of the charging system ${\rho _{AB}}\left( \infty  \right)$ can be described as 
\begin{widetext}
\begin{align}
{\rho _{AB}}\left( \infty  \right)\left| {_{T = {\Delta _A} = {\Delta _B} = 0}} \right. = \frac{1}{{{k^4} + 2\left( {2 + {k^2}} \right){l^2}}}\left( {\begin{array}{*{20}{c}}
  {\frac{{{k^4}}}{4}}&0&{ - \frac{{i{k^3}l}}{2}}&{i{k^2}l} \\
  0&{\frac{{{k^4}}}{4}}&0&{ - \frac{{i{k^3}l}}{2}} \\
  {\frac{{i{k^3}l}}{2}}&0&{\frac{{{k^4} + 4{k^2}{l^2}}}{4}}&{ - 2k{l^2}} \\
  { - i{k^2}l}&{\frac{{i{k^3}l}}{2}}&{ - 2k{l^2}}&{\frac{{{k^4} + 4\left( {4 + {k^2}} \right){l^2}}}{4}}
\end{array}} \right),
\label{Eq.15}
\end{align}
\end{widetext}
where  $k = F/g$, $l = J/g$ and $i$ is an imaginary unit. As a result, the analytical solutions for energy and conversion efficiency in the resonance case can be obtained by combining Eqs. (\ref{Eq.6})-(\ref{Eq.10}) and (\ref{Eq.15}).
\begin{figure}[htbp]
		\centering
		\subfigure{\includegraphics[width=4.2cm]{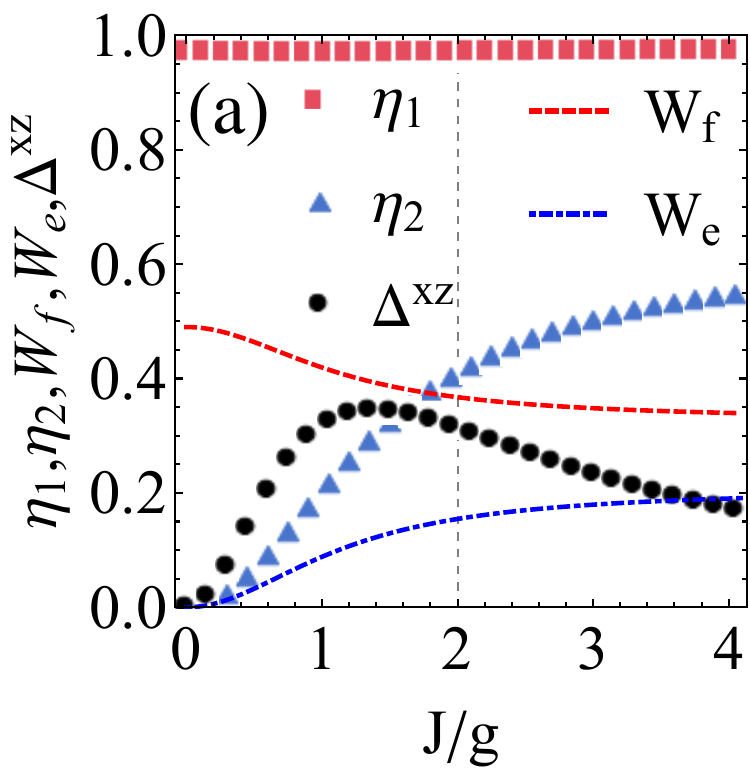}}
		\hspace{0.1cm}
		\subfigure{\includegraphics[width=4.2cm]{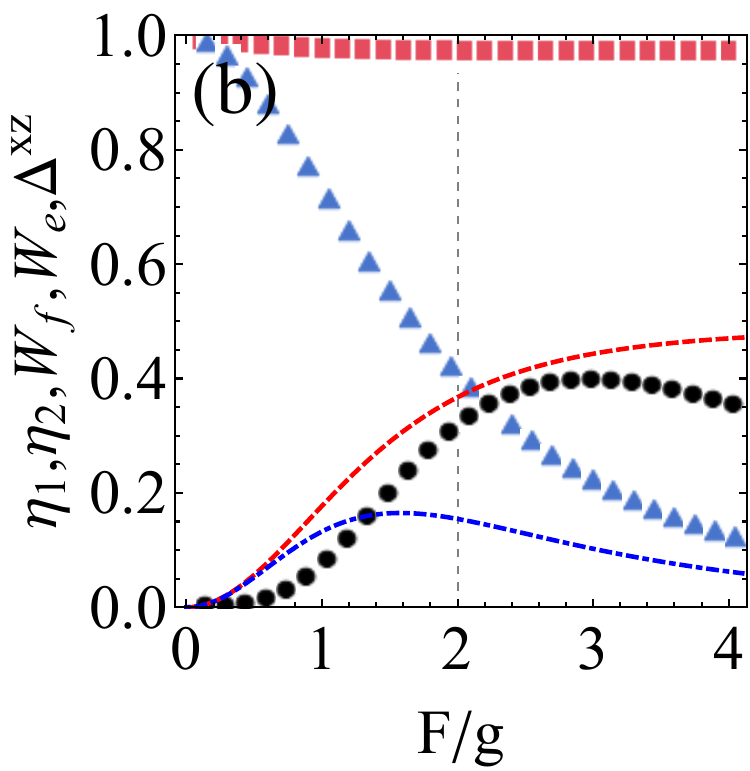}}
		\caption{The extractable work, efficiency and tightness as a function of $J/g$ or $F/g$ for the steady state, where exergy ${{W_f \left( \infty \right)}}$ (the dashed red line), ergotropy ${{W_e \left( \infty \right)}}$ (the dot-dashed blue line), efficiency ${\eta _1}\left( \infty  \right)$ (the red rectangle), efficiency ${\eta _2}\left( \infty  \right)$ (the blue triangle) and bound's tightness $\Delta^{xz}(\infty)$ (black solid circle).  Graphs (a): $F = 2g$. Graphs (b): $J = 2g$. And the thin dashed gray line represents $J/g = F/g$. So $J/g < F/g$ refers to the left area of Graphs (a) and the right area of Graphs (b); $J/g > F/g$ refers to the right area of Graphs (a) and the left area of Graphs (b). All is plotted with $\beta  = 100/{\omega_0}$ and $T ={\Delta _A} = {\Delta _B} = 0$.}
		\label{f2}
	\end{figure}	

Fig. \ref{f2}(a) has plotted the extractable work and efficiency of battery with different $J/g$. Following the figure, one can obtain: the greater $J/g$ is in favor of  ergotropy and ${\eta _2}\left( \infty  \right)$, but against exergy. That is, more ergotropy can be obtained while reducing the cost of charging. In addition, ${\eta _1}\left( \infty  \right)$ has been stable at a high level without being disturbed by $J/g$, i.e., ${{W_f \left( \infty \right)}} \approx {{\Delta {E_B}\left( \infty \right)}}$, which suggests exergy represents the variation of stored energy, and almost all of the stored energy can be extracted into exergy.
Specifically, it can be seen that strong tightness (i.e., the decrease of ${\Delta ^{xz}}\left( \infty  \right)$) can enhance exergy (for $J/g<F/g$) or enhance ergotropy (for $J/g>F/g$) to a certain extent.

Fig. \ref{f2}(b) examines the extractable work and efficiency for different $F/g$. It is found that, 
the extractable work may increase (e.g., exergy) or decreases (e.g., ergotropy) with the increase of $F/g$, where both the efficiency is decreasing, which means a stronger drive will increase the cost of charging. In addition, the strong tightness can increase the extractable work within a certain range.

More importantly, both of Figs. \ref{f2}(a) and \ref{f2}(b)  indicate that strong tightness always corresponds to higher energy conversion efficiency when $J/g>F/g$. To prove generality, this result is fully validated in Fig. \ref{s1} of the \hyperlink{A}{Appendix A}, which proves the result is hold for large-scale range of $J/g\in [0,1000]$. And the strong tightness stems from the fact that the uncertainty continues to decline (i.e., $U_l^{xz}\left( \infty  \right)$ is decreasing) when $J/g>F/g$ (see Fig. \ref{s1}).

In short, when the resonant drive and zero-temperature reservoir co-exist, it is concluded that: (i) the larger dissipation can enhance the extractable work at a lower energy cost (e.g., ergotropy), while a stronger drive cannot; (ii) The variation in tightness always indicates the energy conversion efficiency with respect to $J/g>F/g$.

\begin{figure}[htbp]
			\centering
		\subfigure{\includegraphics[width=4.2cm]{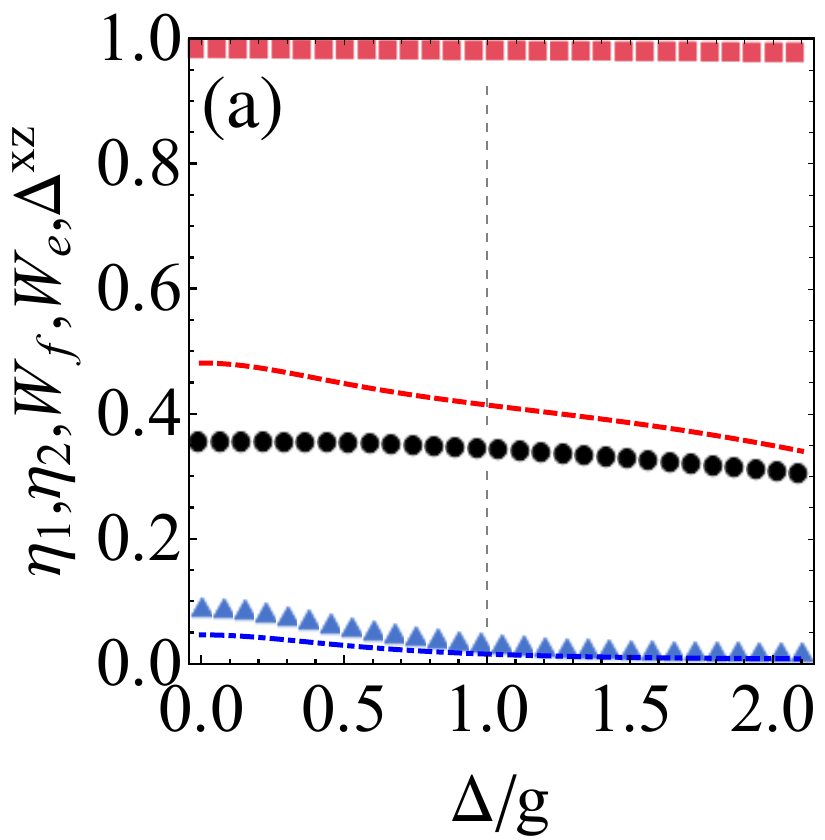}}
		\hspace{0.1cm}
		\subfigure{\includegraphics[width=4.2cm]{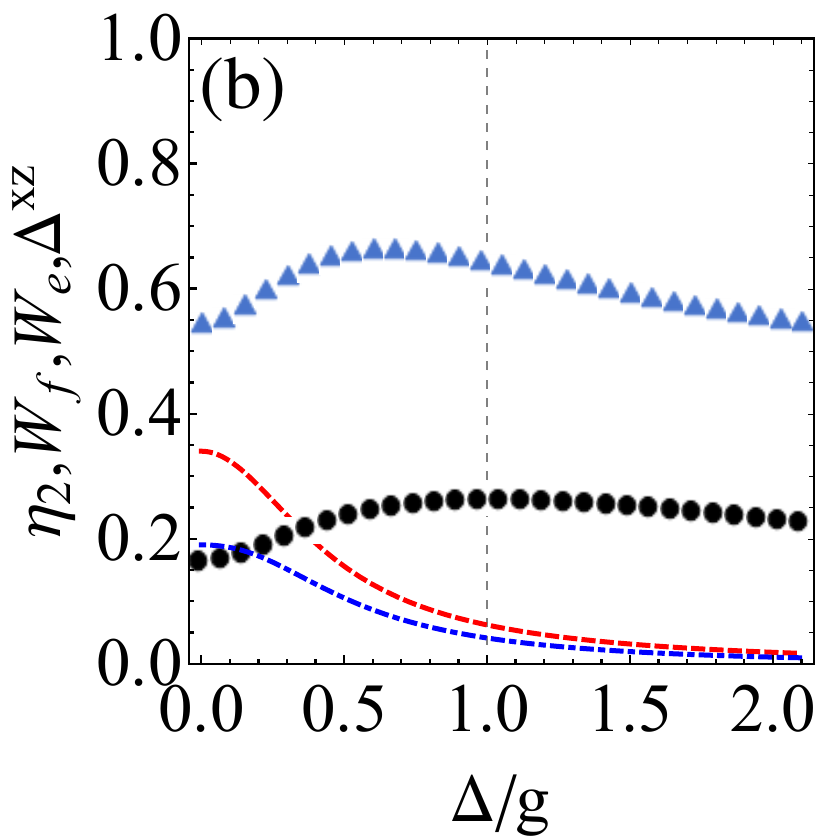}}
		\caption{The extractable work, efficiency and tightness as a function of $\Delta /g$. In addition, we also consider the relative intensity of dissipation and drive, where $F = 2J = 6g$ for Graphs (a) and $J = 2F = 4g$ for Graphs (b). And the thin dashed gray line represents $\Delta = g$.
For all plots, $\beta  = 100/{\omega_0}$, $T  = 0$ and ${\Delta _A} = {\Delta _B} = \Delta $ are set. And all the markers are same as those in Fig. \ref{f2}.}
		\label{f3}
	\end{figure}

\subsection{Detuning}

In this subsection, we further investigate the effect of detuning drive on extractable work and the relationship between tightness and energy conversion efficiency. For simplicity, we assume detuning $\Delta_A=\Delta_B=\Delta$. Therefore, we can attain
the   state   of the charging system ${\rho _{AB}}\left( \infty  \right)\left| {_{T = 0}} \right.$, which
solution is a 4x4 matrix given in the \hyperlink{B}{Appendix B}.

As shown in Fig. \ref{f3}(a), increase of the detuning reduces the extractable work and efficiency ${\eta _2}\left( \infty  \right)$ for $J/g<F/g$. Moreover, the weak tightness (i.e., the increase of ${\Delta ^{xz}}\left( \infty  \right)$) promotes the accumulation of extractable work. Since the efficiency ${\eta _1}\left( \infty  \right)$ is nearly invariable regarding $\Delta/g$, we   will focus on investigating ${\eta _2}\left( \infty  \right)$   in the following.
	
Fig. \ref{f3}(b) illustrates the growth of detuning still inhibits extractable work for $J/g>F/g$. While efficiency may be improved, it comes at the cost of losing extractable work. Combining with  Fig. \ref{f3}(a), detuning is always detrimental to charging of QB. Compared with Fig. \ref{f2}, it finds that the resonant driving field is beneficial to boost the system¡¯s energy,which is a key factor in the course of battery energy storage and extraction. In addition, when $\Delta/g  < 1$, the stronger tightness indicates more extractable work, while the stronger tightness shows less extractable work in the case of $\Delta/g  > 1$.

Furthermore, regardless of the relative strength of dissipation and drive, weak tightness frequently represents a higher energy conversion efficiency. Which suggests that tightness has great potential to indicates the efficiency in detuned charging protocol.

\begin{figure}[htbp]
	\centering
		\subfigure{\includegraphics[width=4.2cm]{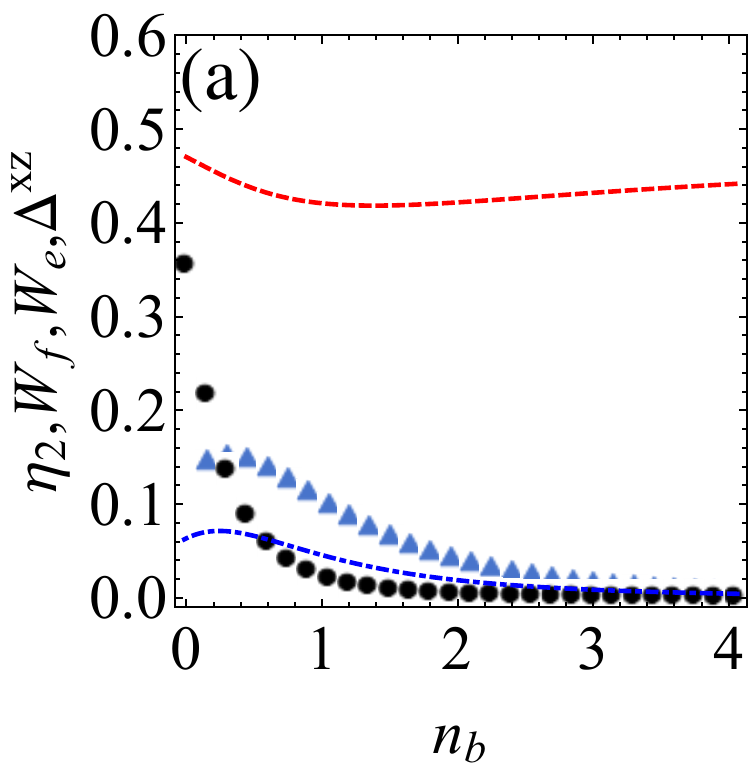}}
		\hspace{0.1cm}
		\subfigure{\includegraphics[width=4.2cm]{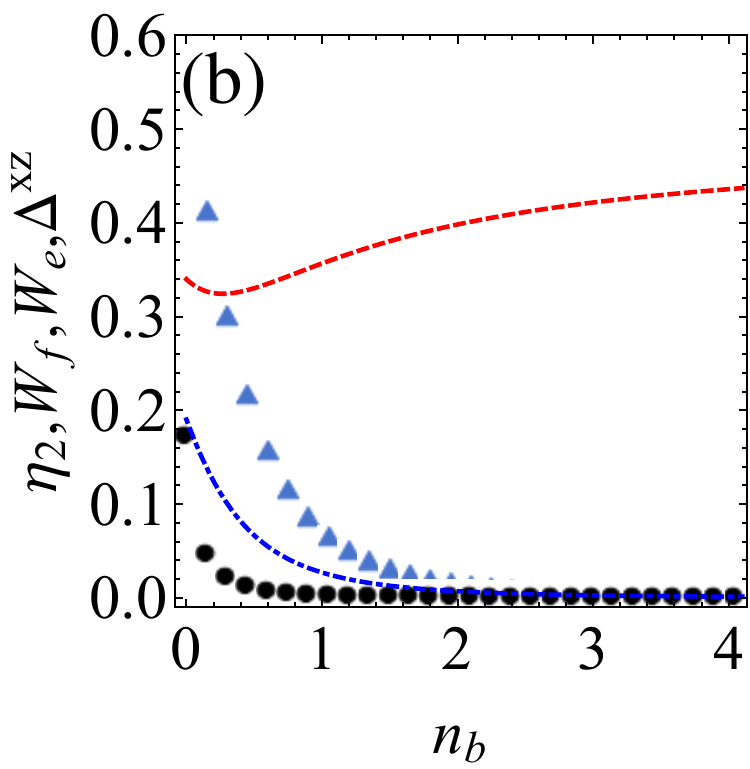}}
		\\
		\subfigure{\includegraphics[width=4.2cm]{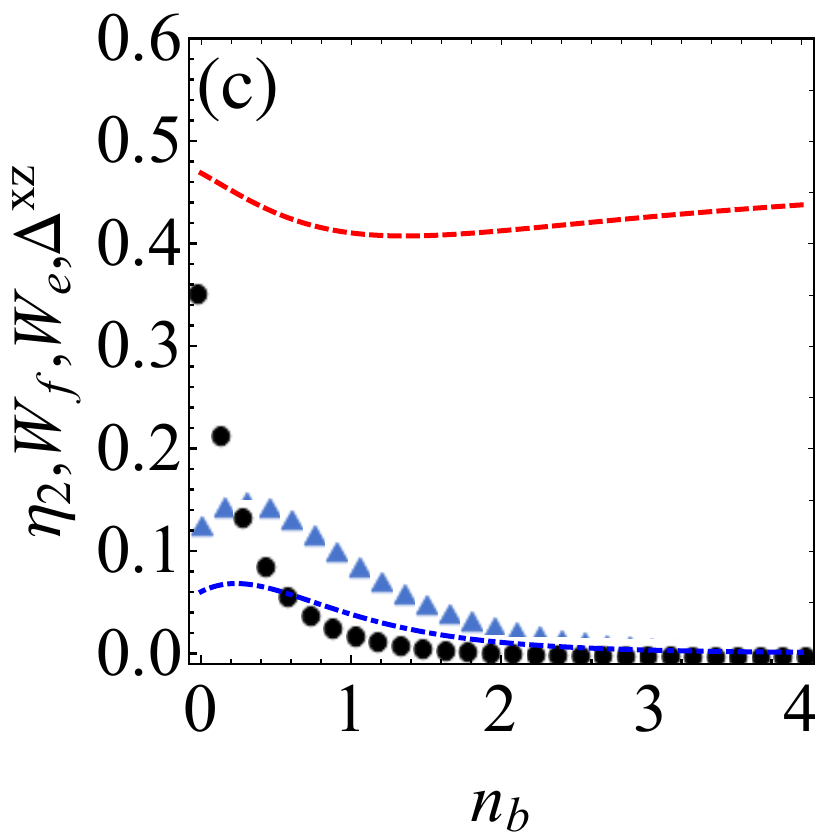}}
\hspace{0.1cm}
\subfigure{\includegraphics[width=4.2cm]{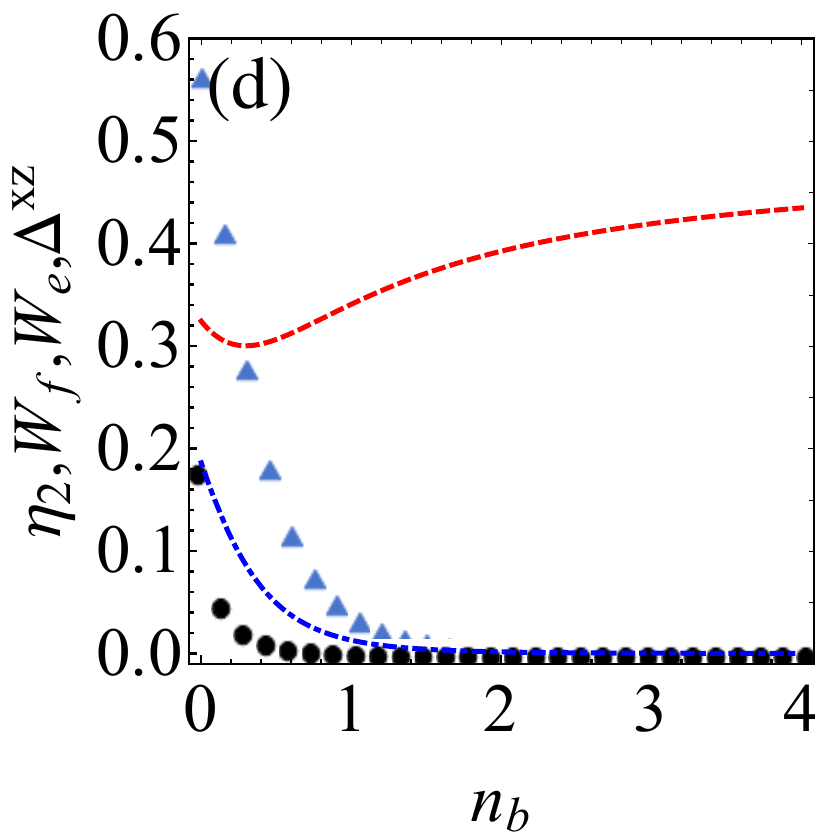}}
		\caption{{For the bosonic reservoir, the extractable work, efficiency and tightness as a function of $n_b$.  Graphs (a) and (c): $F = 2J = 4g$. Graphs (b) and (d): $J = 2F = 4g$. And ${\Delta _A} = {\Delta _B} = 0 $ for (a) and (b); ${\Delta _A} = {\Delta _B} = 0.1g $ for (c) and (d). For all plots, $\beta  = 100/{\omega_0}$ are set. And all the markers are same as those in Fig. \ref{f2}.}}
		\label{f4}
	\end{figure}

\subsection{Bosonic and fermionic reservoirs}

Next, we turn to consider the effects of different reservoirs on extractable work and conversion efficiency. Specifically, the temperature of the reservoirs is in charge of the average particle number. To be explicit, a higher temperature generally leads to an increase in the average particle number. As a matter of fact, it is largely relevant to battery performance. Herein, we examine the connection between the tightness and energy of the QB system of interest in nonzero-temperature bosonic and fermionic reservoirs.
Regarding bosonic and fermionic reservoirs, the charging system can be depicted as ${\rho _{AB}}\left( \infty  \right)\left| {_{{\Delta _A} = {\Delta _B} = 0}} \right.$ for resonance-driven and non-zero temperature reservoirs and
the solutions have been provided in the \hyperlink{B}{Appendix B}.

Figs. \ref{f4}(a) and  \ref{f4}(b) show the effect of increasing average particle number ${n_b}$ on QB in the bosonic reservoir for resonance drive. It can be seen that regardless of the relative magnitude of dissipation and drive, an initial increase in ${n_b}$ may cause a transient mutation in extractable work or efficiency (e.g., ${n_b} \in \left[ {0,0.4} \right]$ for \ref{f4}(a), ${n_b} \in \left[ {0,0.2} \right]$ for \ref{f4}(b)); However, for a greater range of ${n_b}$, the increasing ${n_b}$ may result in decreased ergotropy, efficiency and improved exergy. This manifests that the different average particle number ${n_b}$
has versatile influence on extractable work and efficiency of QBs.

Generally speaking, the high-performance QB requires high efficiency while retaining as much  extractable work as possible, and detuning is destructive to energy, as mentioned before. Therefore, it is necessary to consider a small detuning case while we observe the non-resonance scenario. Figs. \ref{f4}(c)-(d) show the influence of ${n_b}$ on work and efficiency in the presence of a detuning drive. It can be found that the detuning cannot change the variation trend of extractable work and efficiency compared to the case of resonance. Combining with Figs. \ref{f4}(a)-(d), we have that the non-zero temperature bosonic reservoir ($n_b \neq 0$) may benefit exergy while inhibit ergotropy and efficiency compared with zero temperature ($n_b=0$).

In the meantime, ${\Delta ^{xz}}\left( \infty  \right)$ decreases with the increase of ${n_b}$, which suggest weak tightness often improved efficiency in the non-zero temperature bosonic reservoir.

Figs. \ref{f5}(a)-(d) show the effect of the average particle number ${n_f}$ of the fermionic reservoir on extractable work and efficiency in the presence of resonant or detuning drive. The figures shows that non-zero temperature reservoir (i.e., ${n_f}\neq0$) have the potential to improve extractable work and efficiency compared to zero temperature reservoir (i.e., ${n_f}=0$), since the work and efficiency have larger peaks at non-zero temperature. In addition, the variation of ${n_f}$ leads to a phase transition of battery's energy, and we clearly reveal the existence of phase transition point (i.e., $n_f=0.5$) in the \hyperlink{C}{Appendix C} (see Fig. \ref{s2}).

In brief, in the non-zero temperature fermionic reservoir, we obtain: (i) weak tightness is accompanied by improvements in ergotropy and efficiency; Raising exergy requires either strong tightness (before the phase transition) or weak tightness (after the phase transition). (ii) Tightness can predict the energy phase transition of the QB, e.g., the optimal tightness (${\Delta ^{xz}}\left( \infty  \right)=0$) corresponds to the phase transition point in Fig. \ref{f5}.

\begin{figure}[htbp]
	\centering
		\subfigure{\includegraphics[width=4.1cm]{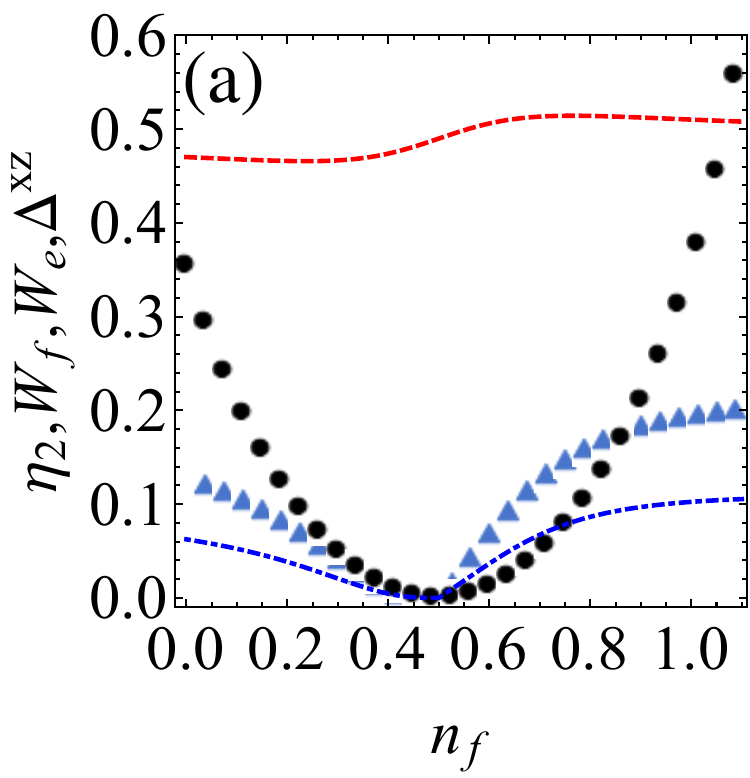}}
		\hspace{0.1cm}
		\subfigure{\includegraphics[width=4.2cm]{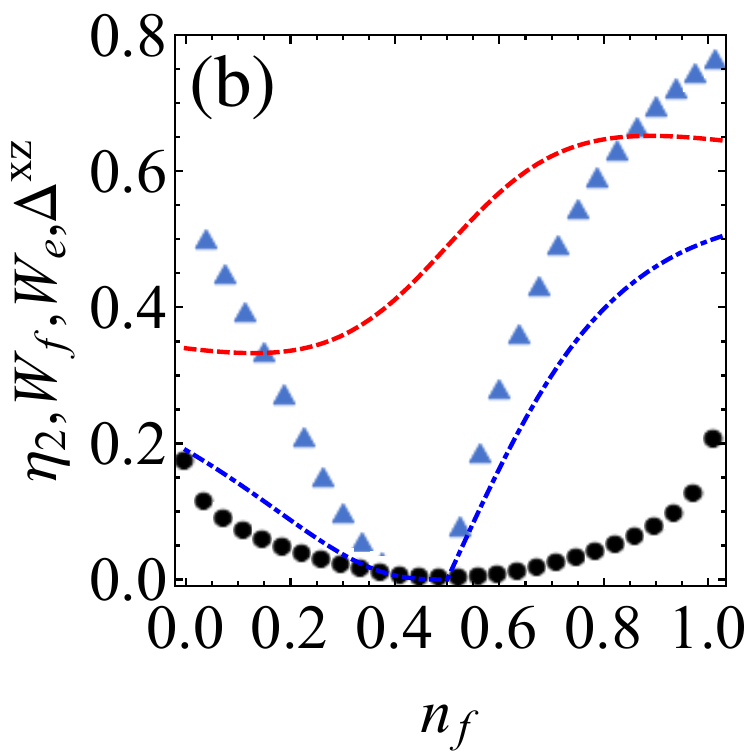}}
		\\
		\subfigure{\includegraphics[width=4.2cm]{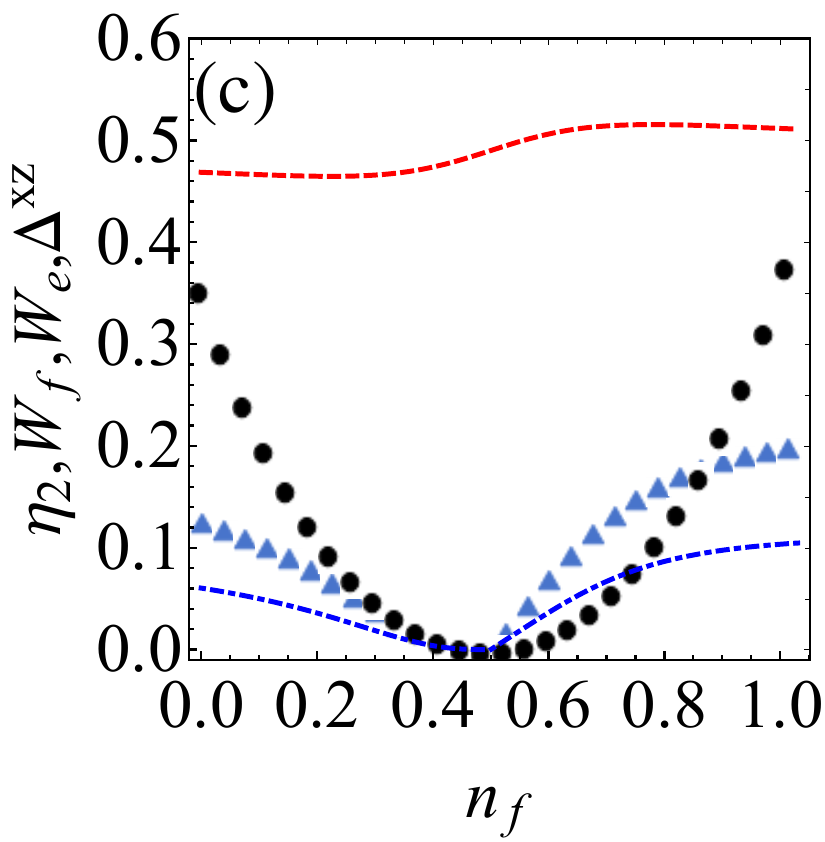}}
\hspace{0.1cm}
\subfigure{\includegraphics[width=4.2cm]{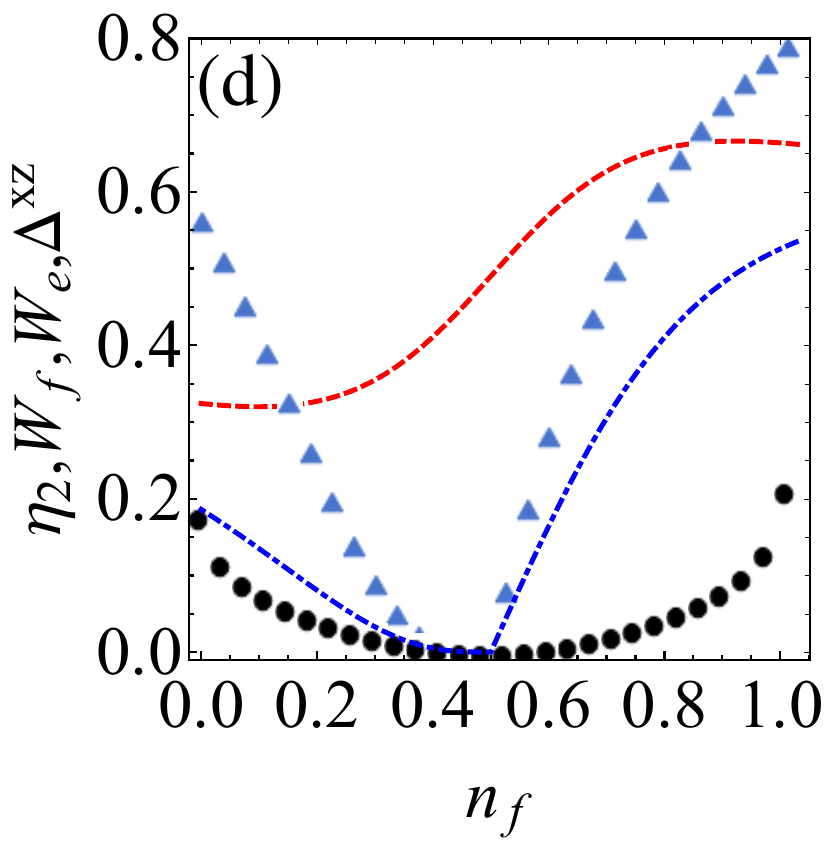}}
		\caption{{For the fermionic reservoir, all settings are same as those in Fig. \ref{f4}.}}
		\label{f5}
	\end{figure}

\section{ Contribution of EUR to the conversion efficiency of QB}
It was found that the tightness of the entropic uncertainty relation was closely associated with the battery's conversion efficiency. However, the tightness is dependent on the uncertainty and its lower bound. Thus, the mechanism by which the EUR affects the efficiency of QB deserves further investigation. Motivated by this, we investigated the dynamics of the lower bound and the uncertainty.

Figs. \ref{f6}(a)-(d) display the dynamics of uncertainty and lower bound. It can be seen that in the case of different parameters regimes, smaller uncertainty and lower bound often represent more efficient energy conversion. In particular, higher efficiency is accompanied by the fact that the distance of the uncertainty from the lower bound (i.e., ${\Delta ^{xz}}\left( \infty  \right)$) is always getting smaller (for Fig. \ref{f6}(a) when $J/g>F/g$) or larger (for Figs. \ref{f6}(b)-(d)). Therefore, the tightness can be an excellent indicator of efficiency.

Owing to the contribution of tightness to efficiency, it concludes that the entropic uncertainty relation is strongly related to the conversion efficiency of extractable work in QBs. The current result provides a new reliable  indicator that predicts energy and efficiency of QBs. \\  \\

\begin{figure}[htbp]
		\centering
		\subfigure{\includegraphics[width=4.2cm]{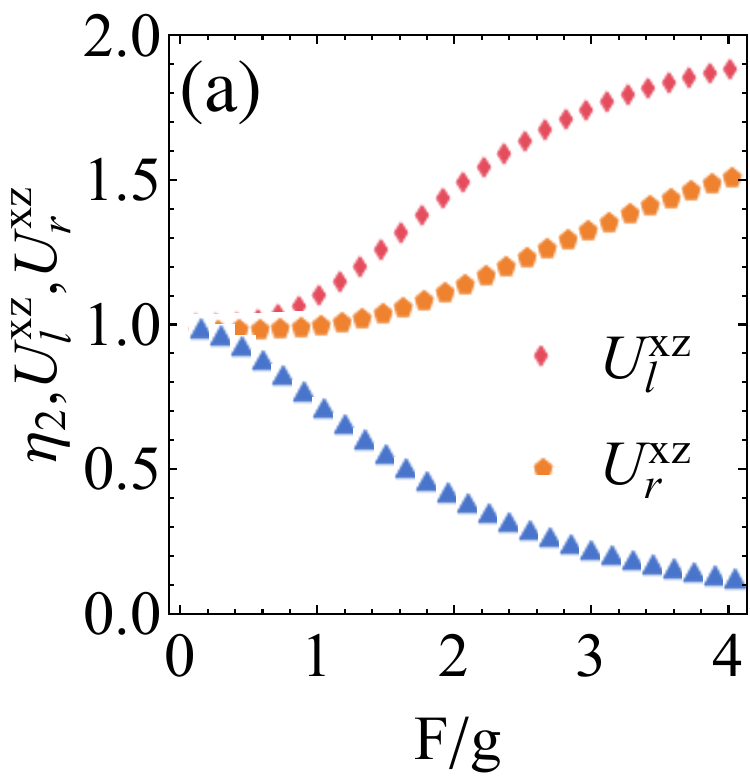}}
		\hspace{0.1cm}
         \subfigure{\includegraphics[width=4.2cm]{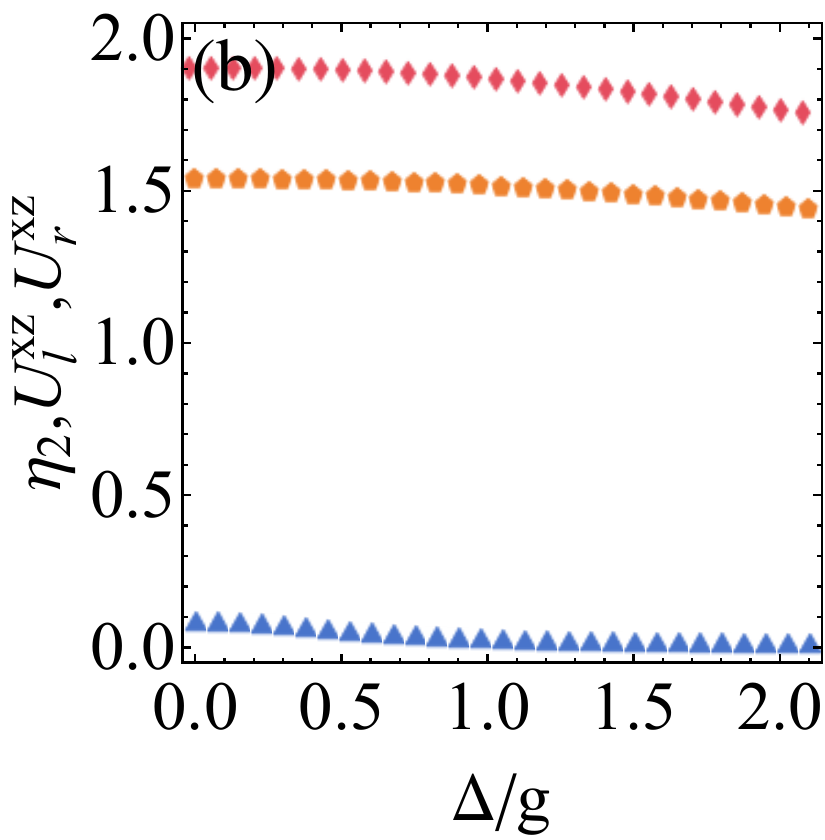}}
		\\
		\subfigure{\includegraphics[width=4.2cm]{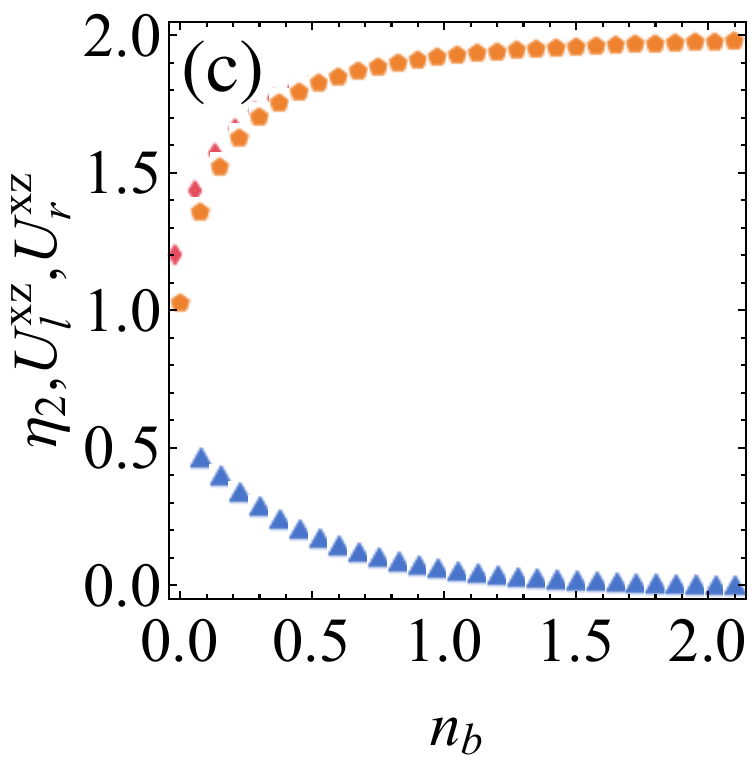}}
		\hspace{0.1cm}
\subfigure{\includegraphics[width=4.2cm]{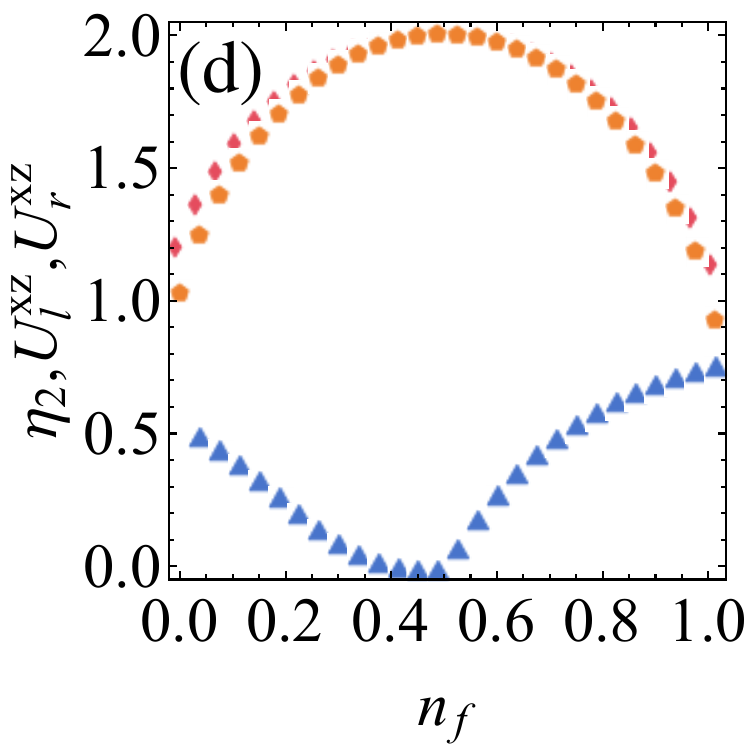}}
		\caption{Uncertainty $U_l^{xz}\left( \infty  \right)$ (red diamond), lower bound $U_r^{xz}\left( \infty  \right)$ (orange pentagon), efficiency ${\eta _2}\left( \infty  \right)$ (blue triangle) as a function of different parameters. Graph (a): $J = 2g$ and  $T ={\Delta _A} = {\Delta _B} = 0$; Graph (b): $F = 2J = 6g$, $T  = 0$ and ${\Delta _A} = {\Delta _B} = \Delta $; Graphs (c) and (d): $J = 2F = 4g$ and ${\Delta _A} = {\Delta _B} = 0 $.  All is plotted for $\beta  = 100/{\omega_0}$.}
		\label{f6}
	\end{figure}

\section{Discussion and conclusion}
In summary, we investigated the extractable work and the conversion efficiency of a QB model with a battery-charger-field in the presence of bosonic or fermionic reservoir using entropic uncertainty relations. We discussed in detail the effect of resonance and detuning between the charger-battery and driving field on the charging process of the QBs and examined how the different reservoirs coupling to the charger affect the extractable work and the efficiency.
Several interesting results can be outlined: (1) resonance drive and higher-temperature reservoirs (both bosonic and fermionic) often induce battery to store more exergy. By contrast, lower-temperature bosonic reservoir favors ergotropy, and more ergotropy leads to more efficiency in most cases. Particularly, stronger dissipation or weaker drive will also improve efficiency. (2) On the one hand, in non-zero temperature reservoirs, strong tightness may increase exergy while suppressing ergotropy. In addition, energy variation induced by dissipation, drive strength and detuning complicates the relationship between tightness and extractable work. On the other hand, the tightness maintains a unified relationship with conversion efficiency compared with extractable work. The stronger tightness always indicates higher efficiency for dissipation greater than drive, when resonant drive and zero-temperature reservoir coexist. Meanwhile the weak tightness is often accompanied by higher efficiency when they do not coexist. (3) In various parametric regimes, smaller uncertainty and its bound often indicate higher efficiency.
Therefore, the current investigations reveal that EURs can be applied to predict the QB's extractable work and its conversion efficiency, which is of fundamental importance in the pursuit of high-performance practical QBs.

\begin{figure}[htbp]
		\centering
		\subfigure{\includegraphics[height=4cm]{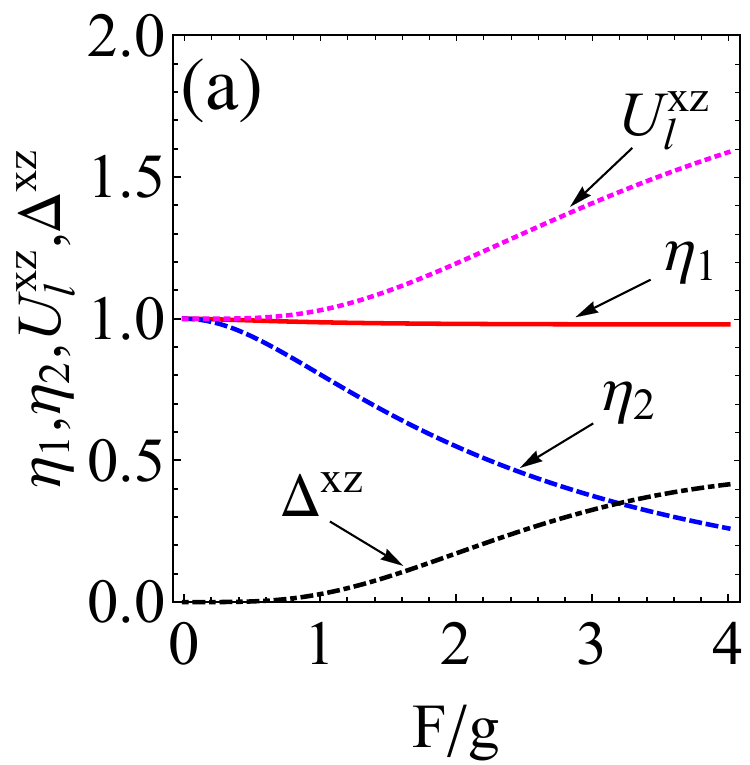}}
		\hspace{0.1cm}
         \subfigure{\includegraphics[height=4cm]{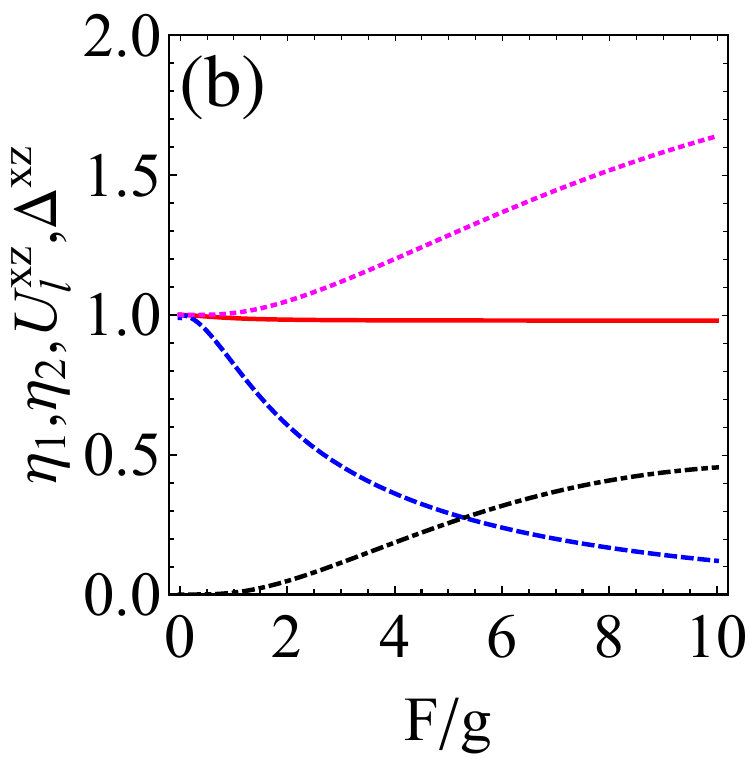}}
		\\
		\subfigure{\includegraphics[height=4cm]{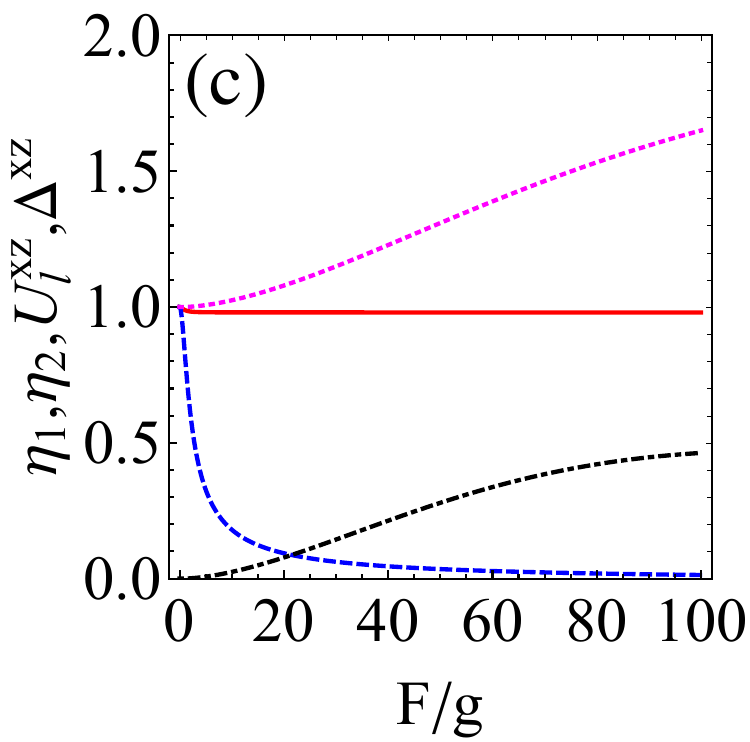}}
		\hspace{0.01cm}
\subfigure{\includegraphics[height=4cm]{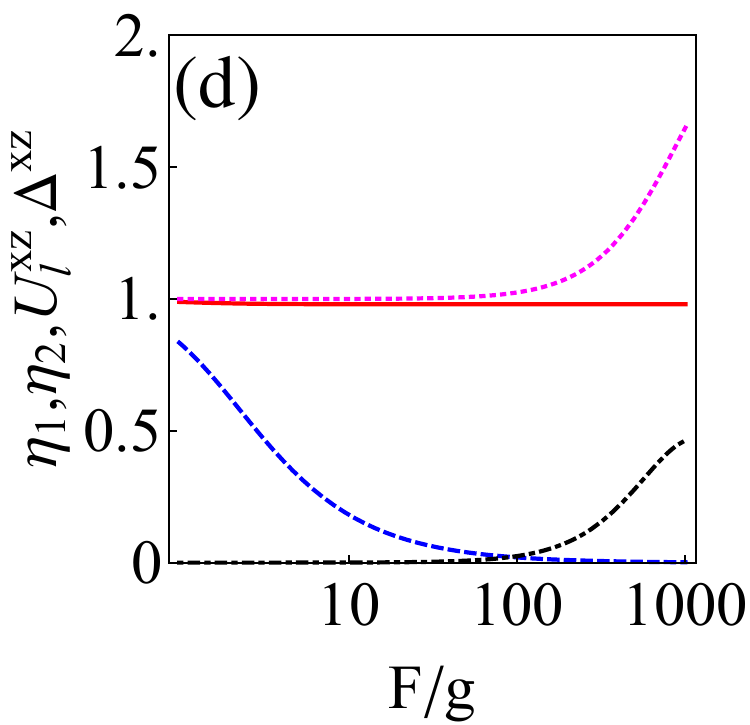}}
		\caption{The efficiency ${\eta _1}\left( \infty  \right)$ (solid red line), ${\eta _2}\left( \infty  \right)$ (dashed blue line), uncertainty $U_l^{xz}\left( \infty  \right)$ (dotted magenta line) and tightness ${\Delta ^{xz}}\left( \infty  \right)$ (dot-dashed black line) as a function of $F/g$ where Graph (a): $J=4g$; Graph (b): $J=10g$; Graph (c): $J=100g$ and Graph (d): $J=1000g$. In addition, the parameters are the same as in Fig. \ref{f2}.}
		\label{s1}
	\end{figure}

	\begin{acknowledgements}
		This study was supported by the National Natural Science Foundation of China (Grant Nos. 12075001, 61601002, 12004006 and 12175001). Anhui Provincial Key Research and Development Plan (Grant No. 2022b13020004), and Anhui Provincial Natural Science Foundation (Grant No. 1508085QF139).
	\end{acknowledgements}

\vskip 3cm

\appendix
\hypertarget{A}{\section{}}

Fig. \ref{f2} has shown that the strong tightness is beneficial for high efficiency of energy conversion in QBs when $J/g>F/g$ and $J/g \in \left[ {0,4} \right]$, and we investigate their relationship over a wider range of $J/g>F/g$ with $J/g \in \left[ {0,1000} \right]$ in Fig. \ref{s1}, which  reveals  the close relationship between tightness and conversion efficiency. That is, in the charging protocol where resonant drive coexist with zero-temperature reservoirs, strong tightness always means higher efficiency if the dissipation of the charger is stronger than the drive field. And the strong tightness is related to the attenuation of uncertainty.

\hypertarget{B}{\section{}}

For the detuning drive and zero-temperature reservoir, the stable charging system is in state of
\begin{align}
{\rho _{AB}}\left( \infty  \right)\left| {_{T = 0}} \right. = \left( {\begin{array}{*{20}{c}}
  {a_{11}}& \cdots &{a_{14}} \\
   \vdots & \ddots & \vdots  \\
  {a_{41}}& \cdots &{a_{44}}
\end{array}} \right),
\label{Eq.16}
\end{align}
where the matrix elements ${a_{ij}}\left( {i,j = 1,2,3,4} \right)$ are given as
\begin{widetext}
\begin{align}
  a_{11} = & ({k^4}({k^4}{m^2} + {k^2}(2 + 4{l^2} + {m^2})(1 + 2{m^2}) + 2({(1 + 2{l^2})^2} + ( - 2 + 5{l^2}){m^2} + {m^4})))/ 4\alpha , \hfill \\
  a_{21} =& \{ ({k^3}m( - 2il{(1 + 2{l^2})^2} + 2{k^2}(2 + {l^2})m + 4{(1 + 2{l^2})^2}m + il(4 + 3{k^2} - 10{l^2}){m^2}\nonumber \hfill \\
   &+ 2( - 4 + {k^2} + 10{l^2}){m^3} - 2il{m^4} + 4{m^5}))\} /2\alpha ,\hfill \\
{a_{31}} = &\{ ({k^3}(2i(1 + {k^2} + 2{l^2})(l + 2{l^3}) - 2(1 + 2{l^2})(2 + {k^2} + 4{l^2})m + il( - 4 + {k^4} + 10{l^2} + {k^2}(5 + 8{l^2})){m^2} \nonumber\hfill \\
   &- ( - 8 + {k^4} + 20{l^2} + {k^2}(2 + 8{l^2})){m^3} + 2i(1 + {k^2})l{m^4} - 2(2 + {k^2}){m^5}))\} / 2\alpha, \hfill \\
{a_{41}} =& \{ ({k^2}( - 4i(1 + {k^2} + 2{l^2})(l + 2{l^3}) + 2(1 + 2{l^2})(4 + 3{k^2} + 10{l^2} + 4{l^4})m \nonumber\hfill \\
   &+ 2il(10 + 3{k^2} - {k^4} + 2(7 - 2{k^2}){l^2} + 24{l^4}){m^2} + 3({k^4} + 4{k^2}{l^2} - 4(2 + {l^4})){m^3} \nonumber\hfill \\
  & + 2il( - 14 + 5{k^2} + 30{l^2}){m^4} - 6( - 4 + {k^2} + 6{l^2}){m^5} + 12il{m^6} - 8{m^7}))\} /2\alpha ,\hfill \\
{a_{12}} =& ({k^3}m(2il{(1 + 2{l^2})^2} + 2{k^2}(2 + {l^2})m + 4{(1 + 2{l^2})^2}m + il( - 4 - 3{k^2} + 10{l^2}){m^2}\nonumber \hfill \\
  & + 2( - 4 + {k^2} + 10{l^2}){m^3} + 2il{m^4} + 4{m^5}))/2\alpha , \hfill \\
{a_{22}} = &\{ ({k^2}({k^6}{m^2} + 8{m^2}({l^2} + 4{m^2})({(1 + 2{l^2})^2} + ( - 2 + 5{l^2}){m^2} + {m^4}) + 2{k^2}({(1 + 2{l^2})^2} + ( - 2 + 9{l^2} + 8{l^4}){m^2}  \nonumber\hfill \\
   &+ 3(3 + 2{l^2}){m^4} + 16{m^6}) + {k^4}(2 + 5{m^2} - 6{m^4} + {l^2}(4 + 8{m^2}))))\} / 4\alpha, \hfill \\
   {a_{32}} = &({k^2}m({k^4}{m^2} + 2{k^2}(1 + 2{l^2} - 7{m^2})(1 + {m^2}) - 4({l^2} + 4{m^2})({(1 + 2{l^2})^2} + ( - 2 + 5{l^2}){m^2} + {m^4})))/ 2\alpha, \hfill \\
{a_{42}} = &\{ (k(i{k^6}(l + im){m^2} + 8(l + 2im)m(il + 2m)(1 + 2{l^2} - 3ilm - {m^2})(1 + 2{l^2} + 3ilm - {m^2})(lm + i( - 1 + {m^2}))\nonumber \hfill \\
   &+ {k^4}(4i{l^3}(1 + 2{m^2}) - 4{l^2}(m + 2{m^3}) + il(2 + 5{m^2} - 6{m^4}) + 2m( - 1 - 5{m^2} + 3{m^4})) \nonumber\hfill \\
  & + 2i{k^2}(8i{l^4}{m^3} + {l^5}(4 + 8{m^2}) + 2i{l^2}m(2 + 4{m^2} + 3{m^4}) + {l^3}(4 + 9{m^2} + 6{m^4}) \nonumber\hfill \\
  & + l(1 - 2{m^2} + 9{m^4} + 16{m^6}) + 2i(m - 6{m^3} - 3{m^5} + 8{m^7}))))\} /2\alpha ,\hfill \\
{a_{13}} = &\{ ({k^3}( - 2i(1 + {k^2} + 2{l^2})(l + 2{l^3}) - 2(1 + 2{l^2})(2 + {k^2} + 4{l^2})m - il( - 4 + {k^4} + 10{l^2} + {k^2}(5 + 8{l^2})){m^2} \nonumber\hfill \\
   &- ( - 8 + {k^4} + 20{l^2} + {k^2}(2 + 8{l^2})){m^3} - 2i(1 + {k^2})l{m^4} - 2(2 + {k^2}){m^5}))\} / 2\alpha,
   \end{align}
\end{widetext}
\begin{widetext}
\begin{align}
{a_{23}} = &({k^2}m({k^4}{m^2} + 2{k^2}(1 + 2{l^2} - 7{m^2})(1 + {m^2}) - 4({l^2} + 4{m^2})({(1 + 2{l^2})^2} + ( - 2 + 5{l^2}){m^2} + {m^4})))/ 2\alpha, \hfill \\
{a_{33}} = &\{ ({k^2}(2(1 + 2{l^2})(1 + {k^2} + 2{l^2})({k^2} + 4{l^2}) + (32 + 20{k^2} + 5{k^4} + {k^6} + 2(56 + 27{k^2} + 6{k^4}){l^2}\nonumber \hfill \\
   &+ 8(21 + 4{k^2}){l^4}){m^2} + 2( - 32 + 3{k^4} + 84{l^2} + {k^2}(9 + 20{l^2})){m^4} + 8(4 + {k^2}){m^6}))\} / 4\alpha, \hfill \\
{a_{43}} = &\{ (k( - 8{l^2}(1 + 2{l^2})(1 + {k^2} + 2{l^2}) - 2i(l + 2{l^3})( - {k^2} + 2(2 + {k^2}){l^2} + 8{l^4})m - 2(16 + 52{l^2} + 68{l^4} \nonumber\hfill \\
   &- 16{l^6} + {k^4}( - 2 + {l^2}) + 10{k^2}(1 + {l^2})){m^2} - il(32 + 20{k^2} - 5{k^4} + 2(56 + {k^2}){l^2} + 168{l^4}){m^3} \nonumber\hfill \\
   & - 2( - 48 + 4{k^2} + 5{k^4} + 4(7 + {k^2}){l^2} - 84{l^4}){m^4} - 2il( - 32 + 15{k^2} + 84{l^2}){m^5} \nonumber\hfill \\
    &+ 4( - 24 + 7{k^2} + 42{l^2}){m^6} - 32il{m^7} + 32{m^8}))\} / 2\alpha, \hfill \\
{a_{14}} =& \{ ({k^2}(4i(1 + {k^2} + 2{l^2})(l + 2{l^3}) + 2(1 + 2{l^2})(4 + 3{k^2} + 10{l^2} + 4{l^4})m + 2il({k^4} + {k^2}( - 3 + 4{l^2})\nonumber \hfill \\
   &- 2(5 + 7{l^2} + 12{l^4})){m^2} + 3({k^4} + 4{k^2}{l^2} - 4(2 + {l^4})){m^3} - 2il( - 14 + 5{k^2} + 30{l^2}){m^4}\nonumber \hfill \\
   &- 6( - 4 + {k^2} + 6{l^2}){m^5} - 12il{m^6} - 8{m^7}))\} /2\alpha , \hfill \\
{a_{24}} =& \{ (k({k^6}( - il - m){m^2} + 8(l - 2im)(l + 2im)m(1 - ilm - {m^2})(1 + 2{l^2} - 3ilm - {m^2})(1 + 2{l^2} + 3ilm - {m^2}) \nonumber\hfill \\
   &+ {k^4}( - 4i{l^3}(1 + 2{m^2}) - 4{l^2}(m + 2{m^3}) + 2m( - 1 - 5{m^2} + 3{m^4}) + il( - 2 - 5{m^2} + 6{m^4}))\nonumber \hfill \\
   &- 2i{k^2}( - 8i{l^4}{m^3} + {l^5}(4 + 8{m^2}) - 2i{l^2}m(2 + 4{m^2} + 3{m^4}) + {l^3}(4 + 9{m^2} + 6{m^4}) \nonumber\hfill \\
   &+ l(1 - 2{m^2} + 9{m^4} + 16{m^6}) - 2i(m - 6{m^3} - 3{m^5} + 8{m^7}))))\} /2\alpha , \hfill \\
{a_{34}} =& \{ (k( - 8{l^2}(1 + 2{l^2})(1 + {k^2} + 2{l^2}) + 2i(l + 2{l^3})( - {k^2} + 2(2 + {k^2}){l^2} + 8{l^4})m - 2(16 + 52{l^2} + 68{l^4} - 16{l^6} \nonumber\hfill \\
  & + {k^4}( - 2 + {l^2}) + 10{k^2}(1 + {l^2})){m^2} + il(32 + 20{k^2} - 5{k^4} + 2(56 + {k^2}){l^2} + 168{l^4}){m^3} \nonumber\hfill \\
  & - 2( - 48 + 4{k^2} + 5{k^4} + 4(7 + {k^2}){l^2} - 84{l^4}){m^4} + 2il( - 32 + 15{k^2} + 84{l^2}){m^5}\nonumber \hfill \\
   &+ 4( - 24 + 7{k^2} + 42{l^2}){m^6} + 32il{m^7} + 32{m^8}))\} /2\alpha , \hfill \\
{a_{44}} = &\{ (({k^8}{m^2} + 32({l^2} + 4{m^2})(1 + ( - 2 + {l^2}){m^2} + {m^4})({(1 + 2{l^2})^2} + ( - 2 + 5{l^2}){m^2} + {m^4}) + {k^6}(2 + 5{m^2} - 2{m^4}\nonumber \hfill \\
   &+ 4{l^2}(1 + 3{m^2})) + 2{k^4}(1 - 6{m^2} + 49{m^4} + 4{m^6} + 12{l^4}(1 + 2{m^2}) + {l^2}(8 + 31{m^2} + 10{m^4}))\nonumber \hfill \\
  & + 8{k^2}(4{l^6}(1 + 3{m^2}) + 4(3 + 5{m^2}){(m - {m^3})^2} + {l^4}(12 + 13{m^2} + 35{m^4})\nonumber \hfill \\
  & + {l^2}(5 + 13{m^2} + 35{m^4} + 43{m^6}))))\} /4\alpha ,
   \label{Eq.C2}
\end{align}
\end{widetext}
with $\alpha={k^8}{m^2} + {k^6}(1 + 2{l^2})(2 + 5{m^2}) + 8({l^2} + 4{m^2})(1 + ( - 2 + {l^2}){m^2} + {m^4})({(1 + 2{l^2})^2} + ( - 2 + 5{l^2}){m^2} + {m^4})
   + 2{k^4}(1 + 17{m^4} + 6{m^6} + 4{l^4}(2 + 3{m^2}) + 3{l^2}(2 + 6{m^2} + 3{m^4})) + 4{k^2}(4{m^2}{( - 1 + {m^2})^2}(2 + 3{m^2})
   + {l^6}(4 + 8{m^2}) + {l^4}(8 + 19{m^2} + 28{m^4}) + {l^2}(3 + 14{m^2} + 35{m^4} + 32{m^6})))$,  $k = F/g$, $l = J/g$, $m = \Delta /g$, and $i$ is an imaginary unit.

In non-zero temperature reservoirs,  the charging system can be expressed as
\begin{align}
{\rho _{AB}}\left( \infty  \right)\left| {_{{\Delta _A} = {\Delta _B} = 0}} \right. = \left( {\begin{array}{*{20}{c}}
  {b_{11}}& \cdots &{b_{14}} \\
   \vdots & \ddots & \vdots  \\
  {b_{41}}& \cdots &{b_{44}}
\end{array}} \right),
\label{Eq.17}
\end{align}
where, the matrix elements  ${b_{ij}}\left( {i,j = 1,2,3,4} \right)$ of ${\rho _{AB}}\left( \infty  \right)\left| {_{{\Delta _A} = {\Delta _B} = 0}} \right.$ in bosonic and fermionic reservoirs corresponds to
\begin{widetext}
\begin{align}
{b_{11}} = &{k^4}(1 + 2{l^2}) + 4{l^2}{n_b}({k^2}(5 + 2{l^2}) + 2{n_b}(2 + 3{k^2} + 2(2 + {k^2}){l^2} + 16{l^2}{n_b}(1 + {n_b})))/4\beta \nonumber \hfill \\
   = &{k^4}(1 + 2{l^2}) + 4{l^2}{n_f}({k^2}(5 - 2{k^2} + 2{l^2}) + 2{n_f}(2 + {k^4} + 4{l^2} - 2{k^2}(1 + 2{l^2}) + 4{k^2}{l^2}{n_f}))/4\gamma , \hfill \\
{b_{21}} =& 2k{l^2}{n_b}( - 1 + 2{l^2} + 4{l^2}{n_b})/ - \beta  \nonumber\hfill \\
   =& 2k{l^2}{n_f}( - 1 + 2{n_f})(1 - 2{l^2} + 4{l^2}{n_f})/ - \gamma ,
 \end{align}
\end{widetext}
\begin{widetext}
\begin{align}
{b_{31}} = &ikl({k^2}(1 + 2{l^2}) + 16{l^2}{n_b}(1 + {n_b}))/2\beta  \nonumber\hfill \\
   = &ikl( - 1 + 2{n_f})({k^2}(1 + 2{l^2}) + 8( - 2 + {k^2}){l^2}( - 1 + {n_f}){n_f})/ - 2\gamma , \hfill \\
{b_{41}} = &i{k^2}l(1 + 2{l^2})/ - \beta \nonumber \hfill \\
   =& i{k^2}l(1 + 2{l^2})/\gamma , \hfill \\
{b_{12}} =& 2k{l^2}{n_b}( - 1 + 2{l^2} + 4{l^2}{n_b})/ - \beta \nonumber  \hfill \\
   =& 2k{l^2}{n_f}( - 1 + 2{n_f})(1 - 2{l^2} + 4{l^2}{n_f})/ - \gamma , \hfill \\
{b_{22}} = &{k^4}(1 + 2{l^2}) + 4{l^2}{n_b}(4 + 5{k^2} + 2(4 + {k^2}){l^2} + 2{n_b}(2 + 3{k^2} + 2(10 + {k^2}){l^2} + 16{l^2}{n_b}(2 + {n_b})))/4\beta \nonumber \hfill \\
   = &{k^4}(1 + 2{l^2}) + 4{l^2}{n_f}(4 + 5{k^2} - 2{k^4} + 2(4 + {k^2}){l^2} + 2{n_f}( - 2 - 2{k^2} + {k^4} - 4(1 + {k^2}){l^2} + 4{k^2}{l^2}{n_f}))/4\gamma , \hfill \\
{b_{32}} =& 0, \hfill \\
{b_{42}} =& ikl({k^2}(1 + 2{l^2}) + 16{l^2}{n_b}(1 + {n_b}))/2\beta  \nonumber \hfill \\
   = &ikl( - 1 + 2{n_f})({k^2}(1 + 2{l^2}) + 8( - 2 + {k^2}){l^2}( - 1 + {n_f}){n_f})/ - 2\gamma, \hfill \\
{b_{13}} = &ikl({k^2}(1 + 2{l^2}) + 16{l^2}{n_b}(1 + {n_b}))/ - 2\beta  \nonumber \hfill \\
   = &ikl( - 1 + 2{n_f})({k^2}(1 + 2{l^2}) + 8( - 2 + {k^2}){l^2}( - 1 + {n_f}){n_f})/2\gamma , \hfill \\
{b_{23}} =& 0, \hfill \\
{b_{33}} =& {k^2}(1 + 2{l^2})({k^2} + 4{l^2}) + 4{l^2}{n_b}(4 + 8{l^2} + {k^2}(7 + 6{l^2}) + 2{n_b}(2 + 3{k^2} + 2(10 + {k^2}){l^2} + 16{l^2}{n_b}(2 + {n_b})))/4\beta  \nonumber \hfill \\
   = &{k^2}(1 + 2{l^2})({k^2} + 4{l^2}) + 4{l^2}{n_f}(4 - 2{k^4} + 8{l^2} + {k^2}(3 - 10{l^2}) + 2{n_f}( - 2 + {k^4} - 4{l^2} + {k^2}( - 2 + 8{l^2}) - 4{k^2}{l^2}{n_f}))/4\gamma , \hfill \\
{b_{43}} =& 2k{l^2}(1 + {n_b})(1 + 2{l^2} + 4{l^2}{n_b})/ - \beta  \nonumber \hfill \\
   =& 2k{l^2}( - 1 + {n_f})( - 1 + 2{n_f})( - 1 - 2{l^2} + 4{l^2}{n_f})/\gamma , \hfill \\
{b_{14}} = &i{k^2}l(1 + 2{l^2})/\beta \nonumber \hfill \\
   =& i{k^2}l( - 1 + 2{n_f})(1 + 2{l^2} + 8{l^2}( - 1 + {n_f}){n_f})/ - \gamma , \hfill \\
{b_{24}} = &ikl({k^2}(1 + 2{l^2}) + 16{l^2}{n_b}(1 + {n_b}))/ - 2\beta \nonumber \hfill \\
   =& ikl( - 1 + 2{n_f})({k^2}(1 + 2{l^2}) + 8( - 2 + {k^2}){l^2}( - 1 + {n_f}){n_f})/2\gamma , \hfill \\
{b_{34}} =& 2k{l^2}(1 + {n_b})(1 + 2{l^2} + 4{l^2}{n_b})/ - \beta \nonumber \hfill \\
   = &2k{l^2}( - 1 + {n_f})( - 1 + 2{n_f})( - 1 - 2{l^2} + 4{l^2}{n_f})/\gamma , \hfill \\
{b_{44}} = &(1 + 2{l^2})({k^4} + 4(4 + {k^2}){l^2}) + 4{l^2}{n_b}(8 + 7{k^2} + 6(8 + {k^2}){l^2} + 2{n_b}(2 + 3{k^2} + 2(26 + {k^2}){l^2} + 16{l^2}{n_b}(3 + {n_b})))/4\beta \nonumber \hfill \\
   =& (1 + 2{l^2})({k^4} + 4(4 + {k^2}){l^2}) + 4{l^2}{n_f}( - 8 + 3{k^2} - 2{k^4} - 2(8 + 5{k^2}){l^2} + 2{n_f}(2 + {k^4} + 4{l^2} + {k^2}( - 2 + 8{l^2}) - 4{k^2}{l^2}{n_f}))/4\gamma,
   \end{align}
\end{widetext}
with $\beta  = (1 + 2{l^2})({k^4} + 2(2 + {k^2}){l^2}) + 8{l^2}{n_b}(1 + {n_b})(2 + 3{k^2} + 2(4 + {k^2}){l^2} + 16{l^2}{n_b}(1 + {n_b}))$,
$\gamma  = (1 + 2{l^2})({k^4} + 2(2 + {k^2}){l^2}) + 8{k^2}{l^2}( - 2 + {k^2} + 2{l^2})( - 1 + {n_f}){n_f}$, and $ {n_b} $ ($ {n_f} $) being the average particle number in the bosonic (fermionic) reservoir.

\begin{figure}[htbp]
		\centering
		\subfigure{\includegraphics[height=4cm]{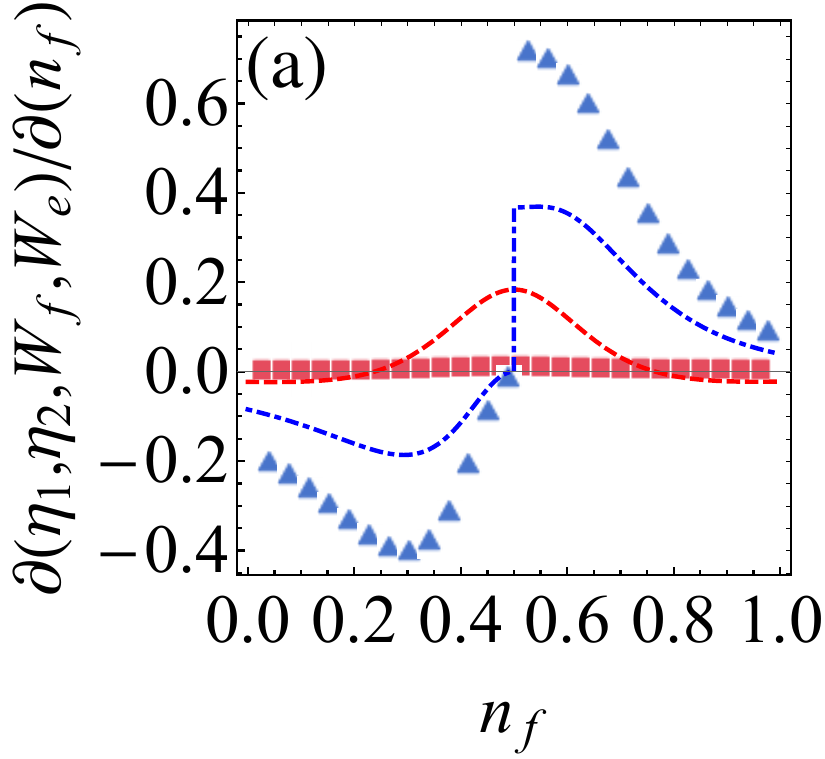}}
		\hspace{0.01cm}
         \subfigure{\includegraphics[height=3.95cm]{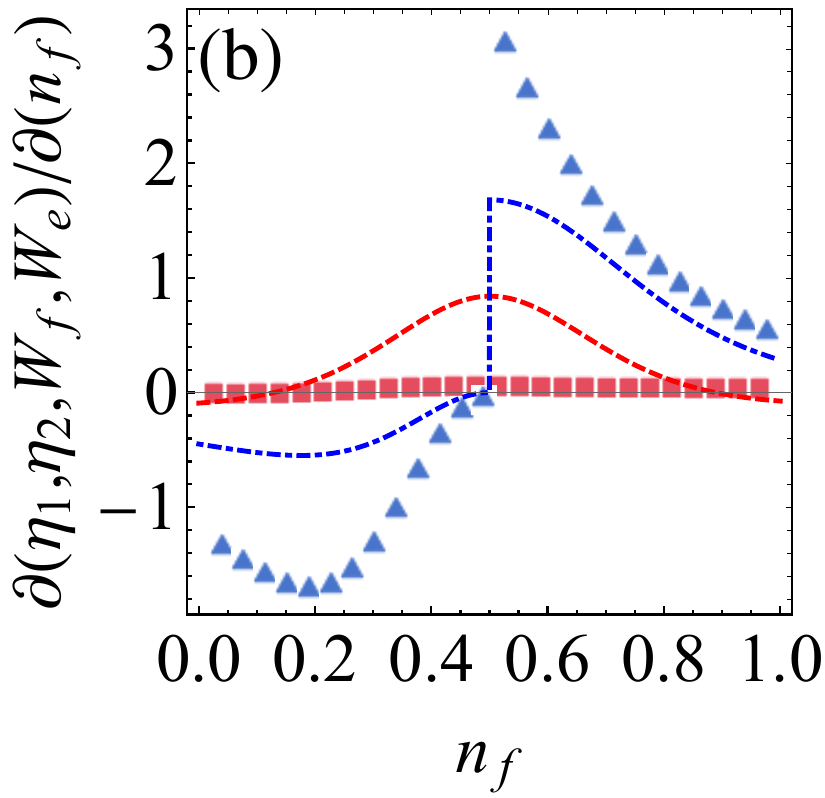}}
		\caption{The first-order partial derivatives of extractable work and efficiency as a function of $n_f$, which includes $\partial {W_f}/\partial {n_f}$ (dashed red line), $\partial {W_e}/\partial {n_f}$ (dot-dashed blue line), $\partial {\eta _1}/\partial {n_f}$ (red rectangle) and $\partial {\eta _2}/\partial {n_f}$ (blue triangle). Graphs (a): $F = 2J = 4g$; Graphs (b): $J = 2F = 4g$. For all plots, ${\Delta _A} = {\Delta _B} = 0 $ and $\beta  = 100/{\omega_0}$ are set.}
		\label{s2}
	\end{figure}

\hskip3cm

\hypertarget{C}{\section{}}

Recent studies \cite{phase1,Dou-F-Q2,cells2} have shown that energy phase transitions play an important role in different charging schemes, and the energy phase transitions that may occur in batteries lead to drastic changes in energy. Quantum phase transition can even lead to adiabatic excitation that increases the energy of the system \cite{phase2}. Therefore, it is necessary to investigate the possible phase transition points in the charging protocol. Shown as Fig. \ref{f5}, in a fermionic reservoir, the changing  average particle number results in a continuous but non-differentiable singularity in the battery's energy.  Fig. \ref{s2} has  depicted the existence of this singularity as a phase transition point (for ${W_e}\left( \infty  \right)$ and ${\eta _2}\left( \infty  \right)$) by calculating the first-order partial derivatives of extractable work and efficiency.

Figs. \ref{s2}(a)-(b) directly show  the presence of phase transition point (i.e., $n_f=0.5$). The partial derivative reaches its peak after the phase transition, i.e., the extractable work and efficiency surge after the phase transition. In combination with Fig. \ref{f5}, it can be seen that the extractable work and efficiency can be rapidly improved after the phase transition and exceed the peak before the phase transition, which supports the positive role of phase transition in the charging protocol.

\end{document}